\documentclass[aps,prb,twocolumn]{revtex4} %,preprint

\usepackage{graphicx}% Include figure files
\usepackage{dcolumn}% Align table columns on decimal point
\usepackage{bm}% bold math

\newcommand{\comment}[1]{}
\begin{document}
%\preprint{}   % Preprint number in upper right corner
\renewcommand{\theequation}{\arabic{section}.\arabic{equation}}

\title{Quantum Statistical Mechanics. IV.
Non-Equilibrium Probability Operator
and Stochastic, Dissipative Schr\"odinger Equation}

%\author{}
%\thanks{}

\author{Phil Attard}
%\affiliation{\protect\texttt{phil.attard1@gmail.com}}

\date{20 June, 2014. phil.attard1@gmail.com}

\begin{abstract}
The  probability operator for a generic non-equilibrium
quantum system is derived.
The corresponding stochastic, dissipative Schr\"odinger equation
is also given.
The dissipative and stochastic propagators are linked
by the fluctuation-dissipation theorem
that is derived from the unitary condition on the time propagator.
The dissipative propagator is derived from
thermodynamic force and entropy fluctuation operators
that are in general non-linear.
\end{abstract}

\pacs{}
%\keywords{}

\maketitle

%%%%%%%%%%%%%%%%%%%%%%%%%%%%%%%%%%%%%%%%%%%%%%%%%%%%%%%%%%%%%%%%%%%%%%%%%%%%%%%
%                                                                             %
                \section*{Introduction}
%                                                                             %
%%%%%%%%%%%%%%%%%%%%%%%%%%%%%%%%%%%%%%%%%%%%%%%%%%%%%%%%%%%%%%%%%%%%%%%%%%%%%%%

What is the probability operator for a non-equilibrium quantum system?

This paper develops the formal theory of
non-equilibrium quantum statistical mechanics
in terms analogous to the author's theory
for the classical non-equilibrium case,
\cite{AttardV,AttardIX,NETDSM,Attard14}
which was itself built on an earlier
approach to classical equilibrium theory.
\cite{Attard00,TDSM}
The present work uses  recent analysis
of equilibrium quantum statistical mechanics,\cite{QSM1,QSM2,QSM3}
which in general this reproduces conventional  text book results,
\cite{Messiah61,Merzbacher70,Bogulbov82}
but with the theory formulated
in terms of the wave function and wave space.\cite{QSM1}
The main new result of that work
was the derivation of the stochastic, dissipative
Schr\"odinger equation for an open equilibrium quantum system.\cite{QSM2}

In order to demonstrate clearly the gap in the current state of knowledge
that the present paper fills,
consider the simplest non-equilibrium system,
namely one in which the energy operator on the sub-system
is time-dependent, $\hat{\cal H}(t)$,
with the sub-system being open and able to exchange
energy with a thermal reservoir of temperature $T$.
For such a mechanical non-equilibrium system,
the non-equilibrium probability operator is not simply
the equilibrium Maxwell-Boltzmann  probability operator
evaluated at time $t$,
\[
\hat \wp(t) \ne \frac{1}{Z(t,T)} e ^{-\hat {\cal H}(t)/k_\mathrm{B}T} ,
\]
where $k_\mathrm{B}$ is Boltzmann's constant.
Almost all books and papers that invoke a time-dependent potential
assert or assume that
this is the non-equilibrium probability operator.
However, it is easy to show that this is not correct.

The second law of thermodynamics
says that for a non-equilibrium system, the entropy must increase
in the positive time direction.
In consequence, the  non-equilibrium probability operator
must possess a time asymmetry that distinguishes
between the future and the past.
Since complex conjugation of an operator corresponds
to velocity reversal,
this means that
%the second law of thermodynamics demands that
the  non-equilibrium probability operator must be complex,
$ \hat \wp(t)^*  \ne \hat \wp(t) $.
But since the Hamiltonian operator is real,
$\hat {\cal H}(t)^* = \hat {\cal H}(t)$,
this proves that
the Maxwell-Boltzmann  probability operator
cannot be the probability operator for a non-equilibrium system.

The correct  non-equilibrium probability operator is herein derived
for generic mechanical and  thermodynamic non-equilibrium  systems.
The detailed physical justification and motivation
for the various definitions that follow
may be found  in Ch.~8 of Ref.~[\onlinecite{NETDSM}].
The present quantum derivation follows
the recently simplified classical version.\cite{Attard14}

A second contribution of the present paper
to non-equilibrium quantum statistical mechanics
is the derivation of the stochastic, dissipative Schr\"odinger
equation for open non-equilibrium systems.
It turns out that this derivation involves non-linear quantum operators.

Research on non-linear operators in quantum mechanics
may be grouped into three main themes:
the formulation of non-linear quantum mechanics,
\cite{Birula76,Weinberg89,Doebner92,Doebner94,Doebner95,Bugajski91,Beretta87}
the non-linear Schr\"odinger equation,
\cite{deBroglie60,Laurent65,Shapiro73,Marinov74,Kupczynski74,%
Mielnik74,Pearle76,Shimony79,Kibble78,Kibble80}
and the non-linear Schr\"odinger equation
including dissipation or friction.
\cite{Messer78,Moxnes05,Lange85,Garashchuk13,Kostin72,Doebner99,Weiss93,%
Dekker81,Schuch12}
In many cases the type of non-linearity is simply postulated
and the emphasis is on the consequences of the chosen form.
In the present paper the non-linear thermodynamic force operator
arises in the first principles derivation and
there is no ambiguity about its final form or specific role
in the dissipative Schr\"odinger equation.
Although this aspect of the present work
belongs to the third category just mentioned,
detailed comparison will not be  made here
because non-linearity \emph{per se} is not the primary focus
of the present work, and also because
in the final form the non-linearity reduces to a scale factor
that only weakly influences the results.

%%%%%%%%%%%%%%%%%%%%%%%%%%%%%%%%%%%%%%%%%%%%%%%%%%%%%%%%%%%%%%%%%%%%%%%%%%%%%%%
%                                                                             %
                \section{Reservoir Entropy}
\setcounter{equation}{0}
%                                                                             %
%%%%%%%%%%%%%%%%%%%%%%%%%%%%%%%%%%%%%%%%%%%%%%%%%%%%%%%%%%%%%%%%%%%%%%%%%%%%%%%

%%%%%%%%%%%%%%%%%%%%%%%%%%%%%%%%%%%%%%%%%%%%%%%%%%%%%%%%%%%%%%%%%%%%%%%%%%
\subsection{Trajectory Entropy}

The total system is isolated and
is composed of a sub-system and a reservoir.
The reservoir is also known as the environment, or the thermal bath,
and the sub-system is also known as an open system.
For the mechanical system with time-dependent external potential
acting on the sub-system,
$\hat{\cal H}(t) = \hat{\cal H}_0 + \hat U_\mathrm{ext}(t)$,
the total energy at time $t$
is the sum of that of the sub-system and that of the reservoir
$E_\mathrm{tot}(t) = E_\mathrm{s}(t) + E_\mathrm{r}(t)$.
Since the total system is isolated,
the change in the total energy from some initial time
is the work performed on the sub-system
by the time-dependent external potential.
An expression for this work will be given shortly.

When the sub-system is in the wave state $\psi$, its energy is
\begin{equation} \label{Eq:Es(psi,t)}
E_\mathrm{s}(\psi,t) =
\frac{1}{N(\psi)}
\langle \psi| \hat{\cal H}(t) |\psi \rangle .
%}{\langle \psi|  \psi \rangle} .
%=
%\frac{\underline{\underline {{\cal H}}}(t)
%: \underline \psi^* \underline \psi
%}{\underline \psi^* \cdot \underline \psi} .
\end{equation}
Here and throughout the magnitude of the wave function is
$N(\psi) = \langle \psi| \psi \rangle$.

In this paper the adiabatic rate of change is defined
as the evolution of the sub-system considered as isolated.
An adiabatic sub-system
evolves according to the deterministic Schr\"odinger equation.
In the present paper
the adiabatic rate of change of a quantity is denoted
by an over-dot  and the superscript 0.
Accordingly, the adiabatic rate of change of the  sub-system energy is
\begin{eqnarray}
\dot E^0_\mathrm{s}(\psi,t)
& = &
\frac{-\dot N^0(\psi)}{N(\psi)} E_\mathrm{s}(\psi,t)
+
\frac{1}{N(\psi)} \langle \dot \psi^0 | \hat{\cal H}(t) |\psi \rangle
\nonumber \\ &  & \mbox{ }
+
\frac{1}{N(\psi)} \langle \psi | \hat{\cal H}(t) | \dot \psi^0 \rangle
+
\frac{1}{N(\psi)}
\langle \psi | \partial_t \hat{\cal H}(t)   |  \psi \rangle
\nonumber \\ & = &
\frac{1}{N(\psi)}
\langle \psi | \partial_t \hat{\cal H}(t) |  \psi \rangle ,
\end{eqnarray}
where $ \partial_t \equiv \partial /\partial t$.
The constancy of the magnitude and of the energy
along the adiabatic trajectory are well-known.
\cite{QSM2,Messiah61,Merzbacher70,Bogulbov82}
Consequently all terms vanish  except
the expectation value of the partial time derivative
of the energy operator.
This single non-vanishing term  represents the rate
at which work is done on the sub-system by the external potential.
Although the  \emph{rate} at which work is done on the sub-system
does not depend upon the trajectory
(i.e.\ whether it is adiabatic for a closed system,
or whether it is dissipative and stochastic for an open system),
the \emph{total} work done
depends upon the actual trajectory up to the present wave state.

In general, a trajectory in the open sub-system,
which is not the adiabatic trajectory,
must conserve the magnitude of the wave function,
$N(\psi(t)) = N(\psi(0)) \equiv N_0$, or
\begin{equation}
\dot N(\psi(t))  = 0 .
\end{equation}
For the adiabatic evolution it can be shown directly that
$\dot N^0(\psi(t))=0$.
For an equilibrium open system,
the constancy of the magnitude occurs because
the time propagator is on average unitary as a consequence
of the reduction condition on the transition probability operator
(see \S IIC of Ref.~\onlinecite{QSM3}).
It is also unitary in the non-equilibrium case
(see \S  IV of Ref.~\onlinecite{QSM3} and Eq.~(\ref{Eq:N(psi(t))}) below).

At time $t$,  denote by $\psi(t)$  the current wave state of the sub-system,
and denote by $\psi[t]$ the trajectory  of the sub-system,
which is to say all of the wave states leading up to the current state.
%It will often prove convenient to discretize the trajectory,
%$[t] = \{ t_0, t_1, \ldots , t_M\}$, with $ t_\alpha = \alpha \tau$
%and $ t = t_M = M \tau $.
%In this case one can write $\psi_\alpha \equiv \psi(t_\alpha)$, and similar.
The total work done on the sub-system is a functional of the trajectory.
Unlike the sub-system energy, for example,
which depends solely on the present state of the system,
the work done depends upon how the system got to the present state.
It is given by
\begin{eqnarray} \label{Eq:W[t]}
W(\psi[t])
& = &
\int_0^t \mathrm{d} t' \;
\frac{1}{N(\psi(t'))}
\langle \psi(t') | \partial_{t'} \hat{\cal H}(t') |  \psi(t') \rangle
\nonumber \\ & = &
\int_0^t \mathrm{d} t' \;
\dot E^0_\mathrm{s}(\psi(t'),t') .
\end{eqnarray}

Although the focus at present is on mechanical non-equilibrium systems,
the formalism is designed to apply as well
to thermodynamic  non-equilibrium systems.
The final equality here is readily expressed
in terms of the adiabatic rate of change
of the so-called static part of the reservoir entropy,
and as such it holds also for
thermodynamic  non-equilibrium systems,
as is discussed following Eq.~(\ref{Eq:hatSr(t)}) below.

The total energy at all times
is the sum of that of the sub-system and that of the reservoir
$E_\mathrm{tot}(t) = E_\mathrm{s}(t) + E_\mathrm{r}(t)$,
irrespective of the total wave state or the history of the system.
Because the total system is isolated,
the rate of change of the total energy is equal to the rate at which work
is being done done on the sub-system,
so that one also has
$E_\mathrm{tot}(\psi[t]) = E_\mathrm{tot}(0) + W(\psi[t])$.
Here the dependence on a particular trajectory of the sub-system is shown.
With these, the reservoir energy for a given sub-system trajectory is
\begin{equation} \label{Eq:E_r[t]}
E_\mathrm{r}(\psi[t]) = \mbox{const.} + W(\psi[t])- E _\mathrm{s}(\psi(t),t).
\end{equation}
The constant is just the starting total energy.
This can be neglected here and below since the start of
a long enough trajectory is uncorrelated with the current position.
Note that $\psi[t]$ is the trajectory of the wave function
of the sub-system, so that this result gives the reservoir energy
irrespective of the reservoir wave state or its history.

Contrast this result with the equilibrium case when the energy operator
is not a function of time.
In such a case
$E_\mathrm{r}(\psi) = \mbox{const.}-E _\mathrm{s}(\psi)$.
This only depends upon the current wave state of the sub-system,
not upon its previous history.\cite{QSM1}

Using the standard thermodynamic result that
the inverse temperature is the energy derivative of the entropy,\cite{TDSM}
$T^{-1} = \partial S(E)/\partial E$,
and the fact that the reservoir is infinitely larger than the sub-system,
a Taylor expansion shows that the reservoir entropy for a particular
trajectory of the sub-system  is
\begin{eqnarray} \label{Eq:S_r[t]}
S_\mathrm{r}^{<>}(\psi[t])
& = &
\mbox{const.} + \frac{E_\mathrm{r}(\psi[t])}{T}
\nonumber \\ & = &
\mbox{const.} +  \frac{W(\psi[t])- E _\mathrm{s}(\psi(t),t)}{T} .
\end{eqnarray}
Again this can be contrasted with the equilibrium result of
$S_\mathrm{r}^{<>}(\psi) = \mbox{const.}  - E _\mathrm{s}(\psi)/T$.
In the language of Ref.~[\onlinecite{QSM2}],
this is the expectation trajectory entropy
rather then the actual trajectory entropy,
as is discussed below.

In paper I,\cite{QSM1}
it was shown that wave states of an isolated system are uniformly weighted.
In consequence, the weight of sub-system wave states
is proportional to the total weight of the reservoir states
for a  given sub-system wave state,
which itself is proportional to the  exponential of the reservoir entropy
in the given sub-system wave state.
Hence the reservoir entropy is the same as the total entropy
$ S_\mathrm{tot}(\psi[t]) = S_\mathrm{r}(\psi[t]) $,
and similarly for the operators.
In this paper the notation $S_\mathrm{r}$ rather than $S_\mathrm{tot}$
will be used, and it will as often be called `the entropy'
as `the reservoir entropy'.
In paper III\cite{QSM3} it was established in a generic fashion
that the exponential of the entropy operator is the probability operator.
In the present non-equilibrium case one expects instead
to be dealing with operator functionals, at least initially.

% end comment %%%%%%%%%%%%%%%%%%%%%%%%%%%%%%%%%%%%%%%%%%%%%%%%%%%%%%%%%%%%

Define the expectation functional $ O(\psi[t])$
as an operator that has expectation value
over a trajectory of
\begin{eqnarray}
O(\psi[t])
& \equiv &
\frac{
\langle \psi[t] | \hat O [t] | \psi[t] \rangle
}{\langle \psi[t] | \psi[t] \rangle}
\nonumber \\ & \equiv &
\int_0^t \mathrm{d} t' \;
\frac{
\langle \psi(t') | \hat O(t')|  \psi(t') \rangle
}{\langle \psi(t') |  \psi(t') \rangle}
\nonumber \\ & = &
\frac{1}{N_0} \int_0^t \mathrm{d} t' \;
\langle \psi(t') | \hat O(t')|  \psi(t') \rangle ,
\end{eqnarray}
since $N(\psi(t')) \equiv N_0 $.
The work done over a trajectory is of this form.

The sub-system energy that appears in Eq.~(\ref{Eq:E_r[t]}) \emph{et seq.},
instead of being written in terms of the current wave state,
as in Eq.~(\ref{Eq:Es(psi,t)}),
can be written as an integral of the total differential over the trajectory,
\begin{equation}
E _\mathrm{s}(\psi(t),t) =
E _\mathrm{s}(\psi(0),0) +
\int_0^t \mathrm{d} t' \;
\frac{\mathrm{d} E _\mathrm{s}(\psi(t'),t') }{\mathrm{d} t'} .
\end{equation}
With this and the expression for the work done,
Eq.~(\ref{Eq:W[t]}),
the expectation  entropy for the trajectory,
Eq.~(\ref{Eq:S_r[t]}),
can be written
\begin{eqnarray} \label{S_r(psi[t])}
\lefteqn{
S_\mathrm{r}^{<>}(\psi[t])
} \nonumber \\ & = &
\mbox{const.}' - \frac{1}{T}
\int_0^t \mathrm{d} t' \;
\left\{
\frac{\mathrm{d} E _\mathrm{s}(\psi(t'),t') }{\mathrm{d} t'}
- \dot E^0_\mathrm{s}(\psi(t'),t')
\right\}
\nonumber \\ & = &
\mbox{const.}' - \frac{1}{T}
\int_0^t \mathrm{d} t' \;
\left\{
\frac{-\dot N(\psi(t'))}{N(\psi(t'))} E_\mathrm{s}(\psi(t'),t')
\right. \nonumber \\ &  & \left. \mbox{ }
+
\frac{1}{N(\psi(t'))}
\langle \dot \psi(t') | \hat{\cal H}(t') |\psi(t') \rangle
\right. \nonumber \\ &  & \left. \mbox{ }
+
\frac{1}{N(\psi(t'))}
\langle \psi(t') | \hat{\cal H}(t') | \dot \psi(t') \rangle
\right\}
\nonumber \\ & = &
\mbox{const.}' -
\frac{1}{T N_0 }
\int_0^t \mathrm{d} t' \;
\left\{
\langle \dot \psi(t') | \hat{\cal H}(t') |\psi(t') \rangle
\right. \nonumber \\ &  &\left. \mbox{ }
+
\langle \psi(t') | \hat{\cal H}(t') | \dot \psi(t') \rangle
\right\}
\nonumber \\ & = &
\mbox{const.}' -
\frac{1}{T N_0}
\int_0^t \mathrm{d} t' \;
\left\{
\langle \psi(t') |
\hat \mathrm{d}_{t'}^\dag
\hat{\cal H}(t') |\psi(t') \rangle
\right. \nonumber \\ &  &  \left. \mbox{ }
+
\langle \psi(t') |
\hat{\cal H}(t') \hat \mathrm{d}_{t'}
| \psi(t') \rangle
\right\} .
\end{eqnarray}
The third equality invokes the constancy of the magnitude.
In this form it is clear that only the non-adiabatic terms
contribute,
since  $\hat \mathrm{d}_{t} \psi(t)
\equiv \mathrm{d}\psi(t)/\mathrm{d}t \equiv \dot \psi(t)$
can be replaced by $\dot \psi(t)-\dot \psi^0(t)$
in the penultimate equality
without changing the result.

The physical interpretation and justification of this result
is straightforward.
The change in reservoir entropy is due to the change in reservoir energy,
which is equal and opposite to the reservoir-induced change
in sub-system energy.
The reservoir-induced change in sub-system energy
is equal to the total change in sub-system energy,
less the adiabatic change in sub-system energy.
The latter is just
the work done on the sub-system by the time varying external potential.
One sees therefore that the integrand of the first equality above
is the total rate of change less the adiabatic
rate of change of the sub-system energy, as required.

From the final equality one sees
that the reservoir  entropy trajectory operator at time $t'$ is
\begin{equation} \label{Eq:hat-Sr[t]}
\hat S_\mathrm{r}[t'] \equiv
\frac{-1}{T}
\left\{ \hat \mathrm{d}_{t'}^\dag  \hat{\cal H}(t')
+
\hat{\cal H}(t')\hat \mathrm{d}_{t'}  \right\}
, \;\; t' \le t .
\end{equation}
Even though this is a local function of time
it only has meaning in the context of a time integral over the trajectory.
In order to make this clear,
brackets are used to encase the argument on the left hand side.
This distinguishes it from
the reservoir entropy operator $\hat S_\mathrm{r}(t)$
that is given shortly and that is to be used in conjunction
with an ordinary expectation value at time $t$.
The conjugate time derivative
$\hat \mathrm{d}_{t}^\dag  \equiv
({\mathrm{d}/{\mathrm{d}t})^\dag} $ acts to the left.
This reservoir entropy operator  is clearly Hermitian.
This result will shortly become the analogue
of Eq.~(8.16) of Ref.~[\onlinecite{NETDSM}].

%The reservoir entropy is the same as the total entropy
%because the entropy of the sub-system wave states is zero.\cite{QSM1}
%Often below, the word `entropy' will be used
%instead of `reservoir entropy' or `total entropy'.

What is ultimately required is  $\hat \wp(t)$,
the probability operator at time $t$
for the non-equilibrium system.
How this is derived from the above reservoir entropy operator functional
(equivalently, entropy trajectory operator)
is the subject of \S \ref{Sec:Sr(psi,t)}.
Prior to that, certain properties
of the transition probability operator and time propagator
are established for later use.

%%%%%%%%%%%%%%%%%%%%%%%%%%%%%%%%%%%%%%%%%%%%%%%%%%%%%%%%%%%%%%%%%%%%%%%%%
\subsection{Transition Probability Operator and Propagator}
%\label{Sec:wp2+U}

Let $\hat \wp(t)$ be the non-equilibrium probability operator
at time $t$, which will be given explicitly below,
and let $\hat \wp^{(2)}(t_2,t_1)$
be the  unconditional transition probability operator
for the transition $ \{ \psi_1, t_1 \} \rightarrow \{ \psi_2, t_2 \}$.
The latter
obeys the reduction condition\cite{QSM3}
\begin{equation} \label{Eq:red-wp2}
\mbox{Tr}^{(1)}_1  \hat \wp^{(2)}(t_2,t_1)
=
\hat \wp(t_2) ,
\end{equation}
and, analogously,
$\mbox{Tr}^{(1)}_2 \hat \wp^{(2)}(t_2,t_1) = \hat \wp(t_1)$.
The subscript on the trace indicates which time is summed over.
The superscript on the one time probability operator
is dropped for simplicity,
$\hat \wp(t) \equiv \hat \wp^{(1)}(t)$.

One can define a stochastic dissipative time propagator
that gives the evolution of the wave function
in the open non-equilibrium system,
\begin{equation}
| \psi(t_2) \rangle = \hat {\cal U}(t_2,t_1)  | \psi(t_1) \rangle .
\end{equation}
An explicit expression for the propagator will be obtained below.
If $t_2 < t_1$, this is a backward trajectory,
and if $t_2 > t_1 $, it is a forward trajectory.

The  conditional transition probability operator
is a two-time operator
that can be written   as the composition of the two one-time time propagators,
\cite{QSM3}
\begin{equation}
\hat \wp^{(2),\mathrm{cond}}(t_2,t_1)
=
\left< \left\{
\hat {\cal U}(t_2,t_1)^\dag ,  \hat {\cal U}(t_2,t_1)
\right\} \right>_\mathrm{stoch}  .
\end{equation}
The notation $\langle \ldots \rangle_\mathrm{stoch}$
signifies the average over the stochastic operators in the propagator.
Accordingly
the unconditional transition probability operator
is the composition
of the conditional transition probability operator
and the singlet probability operator
that can be arranged in four ways,
\begin{eqnarray}
\lefteqn{
\hat \wp^{(2)}(t_2,t_1)
} \nonumber \\
& = &
\left< \left\{
\hat {\cal U}(t_2,t_1)^\dag ,\,   \hat {\cal U}(t_2,t_1) \, \hat \wp(t_1)
\right\} \right>_\mathrm{stoch}
\nonumber \\ & = &
\left< \left\{
\hat \wp(t_1) \hat {\cal U}(t_2,t_1)^\dag ,\,   \hat {\cal U}(t_2,t_1)
\right\} \right>_\mathrm{stoch}
\nonumber \\ & = &
\left< \left\{
\hat {\cal U}(t_2,t_1)^\dag ,\,  \hat \wp(t_2) \, \hat {\cal U}(t_2,t_1)
\right\} \right>_\mathrm{stoch}
\nonumber \\ & = &
\left< \left\{
\hat {\cal U}(t_2,t_1)^\dag \hat \wp(t_2) ,\,   \hat {\cal U}(t_2,t_1)
\right\} \right>_\mathrm{stoch} .
\end{eqnarray}
Taking the traces of this and using the reduction condition,
one obtains the stationarity condition and the unitary condition
for the propagator.
Taking the trace over $t_1$ one obtains
\begin{eqnarray}
\hat \wp(t_2) & = &
\left<
\hat{\cal U}(t_2,t_1) \hat \wp(t_1) \, \hat{\cal U}(t_2,t_1)^\dag
\right>_\mathrm{stoch}
\nonumber \\ & = &
\hat \wp(t_2)
\left< \hat{\cal U}(t_2,t_1) \, \hat{\cal U}(t_2,t_1)^\dag
\right>_\mathrm{stoch} ,
\end{eqnarray}
and taking the trace over $t_2$ one obtains
\begin{eqnarray}
\hat \wp(t_1) & = &
\left<
\hat{\cal U}(t_2,t_1)^\dag \hat{\cal U}(t_2,t_1)
\right>_\mathrm{stoch} \hat \wp(t_1)
\nonumber \\ & = &
\left< \hat{\cal U}(t_2,t_1)^\dag  \hat \wp(t_2)\, \hat{\cal U}(t_2,t_1)
\right>_\mathrm{stoch} .
\end{eqnarray}
The two equations not shown in each case are identical to those shown.
Ordinary composition of one-time operators
is indicated by the juxtaposition of operators in these equations
and in those that follow.
The second equality of the first set,
and the first equality of the second set, imply the unitary condition,
\begin{eqnarray} \label{Eq:wp1-uni}
\hat{\mathrm I}
& = &
\left< \hat{\cal U}(t_2,t_1)^\dag \,
\hat{\cal U}(t_2,t_1)\right>_\mathrm{stoch}
\nonumber \\ & = &
\left< \hat{\cal U}(t_2,t_1) \,
\hat{\cal U}(t_2,t_1)^\dag\right>_\mathrm{stoch},
\end{eqnarray}
and this has complex conjugate
\begin{eqnarray}
\hat{\mathrm I}
& = &
\left< \hat{\cal U}(t_2,t_1)^\mathrm{T}\hat{\cal U}(t_2,t_1)^*
\right>_\mathrm{stoch}
\nonumber \\ & = &
\left< \hat{\cal U}(t_2,t_1)^* \hat{\cal U}(t_2,t_1)^\mathrm{T}
\right>_\mathrm{stoch} .
\end{eqnarray}

The  first equality of the first set of equations,
and the  second equality of the second set of equations,
imply  the stationarity condition,
\begin{eqnarray} \label{Eq:wp1-stat}
\hat \wp(t_2)
& = &
\left<
\hat{\cal U}(t_2,t_1) \hat \wp(t_1) \, \hat{\cal U}(t_2,t_1)^\dag
\right>_\mathrm{stoch}
\nonumber \\ & = &
\left<
\hat{\cal U}(t_1,t_2)^\dag \hat \wp(t_1) \, \hat{\cal U}(t_1,t_2)
\right>_\mathrm{stoch}.
\end{eqnarray}
The nomenclature `stationarity' derives from the equilibrium case.
In the present non-equilibrium situation
it might be better to call it the stability condition,
by which is meant that the form derived
for the probability operator as a function of time
has to be consistent with the time evolution of the probability operator
given by this equation.

Now
\begin{equation}
\hat \wp(t)^\dag =  \hat \wp(t)
\mbox{ but }
\hat \wp(t)^* \ne  \hat \wp(t) .
\end{equation}
The right hand side of the  stationarity condition
is obviously Hermitian.
The complex conjugate of the  stationarity condition is
\begin{eqnarray}
\hat \wp(t_2)^*
& = &
\left<
\hat{\cal U}(t_2,t_1)^* \hat \wp(t_1)^* \, \hat{\cal U}(t_2,t_1)^\mathrm{T}
\right>_\mathrm{stoch}
\nonumber \\ & = &
\left<
\hat{\cal U}(t_1,t_2)^\mathrm{T} \hat \wp(t_1)^* \, \hat{\cal U}(t_1,t_2)^*
\right>_\mathrm{stoch}.
\end{eqnarray}

The unitary nature of the time propagator
implies that on average
the magnitude of the wave function is constant on a trajectory.
That is,
\begin{eqnarray} \label{Eq:N(psi(t))}
\lefteqn{
\left<  N(\psi(t'|\psi_0,t_0)) \right>_\mathrm{stoch}
} \nonumber \\
& = &
\left<
\langle  \psi(t'|\psi_0,t_0) \, | \,  \psi(t'|\psi_0,t_0)  \rangle
 \rule{0cm}{.4cm} % strut:
\right>_\mathrm{stoch}
\nonumber \\ & = &
\left< \langle  \psi_0  | \, \hat{{\cal U}}(t',t)^\dag
\hat{{\cal U}}(t',t) \, |  \psi_0  \rangle
\right>_\mathrm{stoch}
\nonumber \\ & = &
N(\psi_0).
\end{eqnarray}
Here and often below
the wave function trajectory starting at $\{\psi_0,t_0\}$
is written $\psi(t'|\psi_0,t_0)$ rather than the simpler $\psi(t')$ used above.

%%%%%%%%%%%%%%%%%%%%%%%%%%%%%%%%%%%%%%%%%%%%%%%%%
\subsubsection{Operator Evolution}

The time development of an an operator can be derived
from the expectation value
\begin{eqnarray}
\lefteqn{
O(\psi(t|\psi_0,t_0),t)
} \nonumber \\
& = &
\frac{1}{N_0} \langle \psi(t|\psi_0,t_0)  | \hat O(t)
| \psi(t|\psi_0,t_0)   \rangle
\nonumber \\ & = &
\frac{1}{N_0} \langle \psi_0 | \hat{{\cal U}}(t,t_0)^\dag
\hat O(t) \hat{{\cal U}}(t,t_0) | \psi_0 \rangle.
\end{eqnarray}
Hence
\begin{equation} \label{Eq:O(t,t0)}
\hat O(t,t_0) =
\left< \hat{{\cal U}}(t,t_0)^\dag \hat O(t) \hat{{\cal U}}(t,t_0)
\right>_\mathrm{stoch}
\end{equation}
is the evolved operator to be used in conjunction with
the original wave state at time $t_0$.
The stochastic average of this has been taken.
Note the similarities and differences between this expression
for the observable operator evolution
and that for the probability  operator evolution,
Eq.~(\ref{Eq:wp1-stat}).

The time derivative of the evolved operator
(with respect to the final state) is
\begin{eqnarray}
\left. \frac{\mathrm{d}\hat O(t,t_0)}{\mathrm{d}t} \right|_{t_0 \rightarrow t}
& \equiv & \hat{\dot O}(t)
 \\ \nonumber & = &
\partial_t \hat O(t)
+ \hat{\dot{\overline{\cal U}}}(t)^\dag \hat O(t)
+ \hat O(t) \hat{\dot{\overline{\cal U}}}(t)
 \\ \nonumber && \mbox{ }
+ \frac{1}{\tau}
\left< \hat{\tilde{\cal R}}(\tau,t)^\dag \hat O(t) \hat{\tilde{\cal R}}(\tau,t)
\right>_\mathrm{stoch} ,
\end{eqnarray}
with $\tau \equiv t-t_0 \rightarrow 0$.
This is the total (or convective, or hydrodynamic) time derivative.
It takes into account the changes in the expectation value of the operator
as the system moves along its trajectory in wave space.
It expresses the propagator as the sum of a most likely part
and a stochastic part of zero mean,
$ \hat{{\cal U}} =  \hat{{\overline{\cal U}}} +  \hat{\tilde{\cal R}}$,
as will be given explicitly later.
Due to irreversibility,
$\hat{\dot{\overline{\cal U}}}(t)$ depends upon the sign of $\tau$,
and one should be careful to specify whether the forward or backward
derivative is required.
%The stochastic average of this would replace the time derivative
%of the propagators by the time derivative
%of their most likely value.
The partial time derivative,
which only accounts for the changes in the operator due
to its explicit time dependence,
is denoted
$ \partial_t \hat O(t) \equiv {\partial \hat O(t)}/{\partial t}$.
The adiabatic  total time derivative of an operator is
\begin{eqnarray}
\hat{\dot O}\,\!^0(t)
& \equiv &
\partial_t \hat O(t)
+ \hat{\dot{{\cal U}}}\,\!^0(t)^\dag \hat O(t)
+ \hat O(t) \hat{\dot{{\cal U}}}\,\!^0(t)
\nonumber \\ & = &
\partial_t \hat O(t)
- \frac{1}{i\hbar} \hat{\cal H}(t) \hat O(t)
+ \frac{1}{i\hbar} \hat O(t)  \hat{\cal H}(t).
\end{eqnarray}

\comment{ %%%%%%%%%%%%%%%%%%%%%%%%%%%%%%%%%%%%%%%%%%%%%%%%%%%%%%%
%%%%%%%%%%%%%%%%%%%%%%%%%%%%%%%%%%%%%%%%%%%%%%%%%%
\subsubsection{Trajectory Conjugate Hamiltonian [notokpa?]}

For the case of a mechanical non-equilibrium system,
with time-dependent $\hat{\cal H}(t)$,
for microscopic reversibility to hold
\emph{all} of the velocities in the universe need to be reversed.
Under such a reversal one obtains the trajectory conjugate Hamiltonian,
which is just a time reversed version of the original.
With the Hamiltonian operator $\hat{\cal H}(t)$
defined on the time interval $[t_1,t_f]$,
then the trajectory conjugate Hamiltonian on the same time interval is
\begin{equation}
\hat{\tilde {\cal H}}(t)
=
\hat{\cal H}(t_1+t_f-t) .
\end{equation}
The original propagator may be denoted $\hat{U}(t_f, t_1)$,
and the propagator corresponding to the conjugate Hamiltonian
may be denoted $\hat{\tilde U}(t_f, t_1)$.
One has
$\langle\hat{\tilde U}(t_f,t_1)^* \,
\hat{U}(t_f, t_1) \rangle_\mathrm{stoch}
= \hat{\mathrm I}$.
The conjugate Hamiltonian is not necessary for  Hermitian conjugation
because one has in this case
$\langle \hat{ U}(t_f, t_1)^\dag \,
\hat{U}(t_f, t_1) \rangle_\mathrm{stoch}
= \hat{\mathrm I}$.
}  % end comment{ %%%%%%%%%%%%%%%%%%%%%%%%%%%%%%%%%%%%%%%%%%%%%%%%%%%%%%%

%%%%%%%%%%%%%%%%%%%%%%%%%%%%%%%%%%%%%%%%%%%%%%%%%%%%%%%%%%%%%%%%%%%%%%%%%%
\subsection{Point Entropy} \label{Sec:Sr(psi,t)}

\subsubsection{Average Propagator Product}

The approach taken in \S 8.2.2 of Ref.~[\onlinecite{NETDSM}]
for obtaining the non-equilibrium probability density
of a classical phase space point
from the non-equilibrium classical trajectory probability
is now converted into quantum terms.
The analogous quantum procedure
would be to go from a trajectory to a wave state
(or from an expectation value functional to an expectation value)
by replacing the arbitrary trajectory $\psi[t]$
by the most likely trajectory that passes through the wave state $\psi$
at time $t$, which may be denoted $\overline \psi(t'|\psi,t)$.
%This is denoted $\overline \psi(t'|\psi,t)$,
%when a particular time on the trajectory is needed,
%or else by $\overline \psi[t'|\psi,t]$
%for the functional of the trajectory.
However, in the quantum case
it turns out that the stochastic average over the trajectories
rather than the most likely trajectory is required.
%(In the classical case these are the same.)

To see why, the most likely trajectory
can be expressed in terms of the most likely time propagator,
\begin{equation}
| \overline \psi(t'|\psi,t) \rangle
= \hat{\overline{\cal U}}(t',t) |  \psi \rangle .
\end{equation}
%The most likely time propagator does not obey
%microscopic reversibility to leading order,
%but it turns out that the magnitude is still conserved
%on the most likely trajectory to leading order,
%$N(\overline \psi(t'|\psi,t)) = N(\psi)$.
%This is discussed in more detail below.
The most likely propagator and the average propagator are equal,
\begin{equation}
\left<  \hat{{\cal U}}(t',t) \right>_\mathrm{stoch}
= \hat{\overline{\cal U}}(t',t)   .
\end{equation}
However, due to correlations,
the average of the product is not equal to the product of the averages,
\begin{equation}
\left<  \hat{{\cal U}}(t',t)^\dag  \,
\hat{{\cal U}}(t',t) \right>_\mathrm{stoch}
\ne
\hat{\overline{\cal U}}(t',t)^\dag \, \hat{\overline{\cal U}}(t',t) .
\end{equation}
(From the above, the left hand side is the unit operator.)
In what follows, the product of the propagators appears,
and it is for this reason the point reservoir entropy
is formulated as the stochastic average of the reservoir trajectory  entropy.

%%%%%%%%%%%%%%%%%%%%%%%%%%%%%%%%%%%%%%%%%%%%%%%%
\subsubsection{Reduction of the Trajectory Entropy} \label{Sec:Red-Traj-S}

The trajectory may be written in terms of the propagator as
$| \psi(t') \rangle
\equiv | \psi(t'|\psi,t) \rangle
\equiv \hat{\cal U}(t',t) | \psi \rangle $.
The reservoir entropy operator
that was to be used in conjunction with a time integral over
the trajectory was given above as Eq.~(\ref{Eq:hat-Sr[t]})
\begin{equation}
\hat S_\mathrm{r}[t'] \equiv
\frac{-1}{T}
\left\{ \hat \mathrm{d}_{t'}^\dag  \hat{\cal H}(t')
+
\hat{\cal H}(t')\hat \mathrm{d}_{t'}  \right\}
, \;\; t' \le t .
\end{equation}
This acts on the trajectory itself, $| \psi[t'|\psi,t] \rangle$.
In view of the propagator form for the trajectory, one may alternatively
define the reservoir entropy operator for the trajectory as
\begin{eqnarray}
\hat S_\mathrm{r}[t';t]  & \equiv &
\hat{\cal U}(t',t)^\dag \, \hat S_\mathrm{r}[t'] \, \hat{\cal U}(t',t)
 \\ \nonumber & = &
\frac{-1}{T}
\hat{\cal U}(t',t)^\dag
\left\{ \hat \mathrm{d}_{t'}^\dag  \hat{\cal H}(t')
+
\hat{\cal H}(t')\hat \mathrm{d}_{t'}  \right\} \hat{\cal U}(t',t) .
\end{eqnarray}
This acts on the current wave function $|\psi\rangle$.

The expectation entropy for the trajectory,
Eq.~(\ref{S_r(psi[t])}),
can be equally well written in terms of either of these,
\begin{eqnarray}
S_\mathrm{r}^{<>}(\psi[t])
& = &
\frac{1}{N_0}
\int_0^t \mathrm{d} t' \;
\langle \psi(t') | \hat S_\mathrm{r}[t']  | \psi(t') \rangle
\nonumber \\ & = &
\frac{1}{N_0}
\int_0^t \mathrm{d} t' \;
\langle \psi  | \hat S_\mathrm{r}[t';t]  | \psi  \rangle ,
\end{eqnarray}
where the magnitude $N_0 \equiv
\langle \psi(t') | \psi(t') \rangle$
is constant on the trajectory.

One may average over the stochastic trajectories
and perform the time integral
to obtain the reservoir entropy operator
for a wave state at time $t$,
\begin{equation} \label{Eq:^Sr(t';t)=>^Sr(t)}
\hat S_\mathrm{r}(t) \equiv
\int_0^t \mathrm{d} t' \;
\left<  \hat S_\mathrm{r}[t';t]
\right>_\mathrm{stoch} .
\end{equation}
This is a single time operator that is evaluated in explicit detail below.
In terms of this the expectation entropy for the current wave state $\psi$ is
\begin{equation}
S_\mathrm{r}^{<>}(\psi,t)
=
\frac{\langle \psi | \hat S_\mathrm{r}(t)  | \psi \rangle
}{\langle \psi | \psi \rangle},
\end{equation}
and the actual entropy is
\begin{equation}
S_\mathrm{r}(\psi,t)
=
k_\mathrm{B} \ln
\frac{\langle \psi | e^{ \hat S_\mathrm{r}(t) /k_\mathrm{B}} | \psi \rangle
}{\langle \psi | \psi \rangle} .
\end{equation}
For the distinction between the expectation entropy and the actual entropy,
see Eqs~(1.30) and (1.31) of Ref.~[\onlinecite{QSM3}];
it is the actual entropy that fundamentally gives the probability density
of the wave state.

Taking the time integral of the stochastic average
of  reservoir entropy trajectory operator
$ \hat S_\mathrm{r}[t';t]$
reduces it to the single time operator $ \hat S_\mathrm{r}(t) $.
The latter is \emph{the} reservoir entropy operator
for the non-equilibrium system.
As such the expression (\ref{Eq:^Sr(t';t)=>^Sr(t)})
is a manifestation of the reduction condition.

The reduction condition\cite{NETDSM,AttardII}
in its original form stated that the second entropy of two states
evaluated at the optimum value of one of the states
is equal to the first entropy of the other state.
The reduction condition is essentially
a statement that fluctuations contribute negligibly to the entropy,
in which case the optimum value is the same as the average value.
Originally derived for classical systems,
the quantum analogue has also been shown to hold.\cite{QSM3}
An important outcome of the quantum analysis
was the distinction between the expectation entropy
and the actual entropy,
and the conclusion that
it was the actual entropy to which the reduction condition applied.\cite{QSM3}

Because of the close relationship between entropy and probability,
the reduction condition on the entropy is similar to,
but not precisely the same as,
the reduction condition on the transition probability,
Eq.~(\ref{Eq:red-wp2}).
The latter involves a sum over all states,
whereas the former takes the logarithm of that sum
and singles out the most likely state as dominant.

The reduction condition can obviously be generalized to larger sets of states.
\cite{NETDSM}
In the case of a trajectory,
it says that if the trajectory is optimized
with respect to all points except the final one,
then the trajectory entropy is equal to the ordinary (first) entropy
of that terminal point.
Equivalently, the stochastic average over all points on the
trajectory except one reduces the trajectory entropy
to the state entropy of the remaining point.

Strictly speaking,
there should be added to the expression for the reservoir entropy,
Eq.~(\ref{Eq:^Sr(t';t)=>^Sr(t)}),
a constant operator in wave space,
namely the identity operator multiplied by a time-dependent scalar
that is the difference between
the current time value of the reservoir entropy
and the running time average of the  reservoir entropy,
\[
\left[ \overline S_\mathrm{r}(t)
- \frac{1}{t} \int_0^t \mathrm {d} t'\,  \overline S_\mathrm{r}(t')
\right] \hat{\mathrm I} .
 \]
This is the analogue of the  scale factors
for the reduction condition on the non-equilibrium weights
(see Eq.~(4.4) of Ref.~[\onlinecite{QSM3}],
and  Eq.~(8.38) of  Ref.~[\onlinecite{NETDSM}]).
Although this constant has a definite physical interpretation,
it can be discarded
(i.e.\ incorporated into the normalizing partition function
for the probability operator)
without affecting any statistical average.
As such, $S_\mathrm{r}(\psi,t)$
is more precisely called the wave state dependent part
of the reservoir entropy at the current time.

%%%%%%%%%%%%%%%%%%%%%%%%%%%%%%%%%%%%%%%%%%%%%%%%
\subsubsection{Form and Interpretation of the Point Entropy}

The reservoir (equivalently total)
entropy operator,
Eq.~(\ref{Eq:^Sr(t';t)=>^Sr(t)}),
in the case of a mechanical non-equilibrium system
may be explicitly written and rearranged as
\begin{eqnarray} \label{Eq:hatSr(t)}
\lefteqn{
\hat S_\mathrm{r}(t)
} \nonumber \\
& = &
\frac{-1}{T }
\int_0^t \mathrm{d} t' \;
%\nonumber \\ &  & \mbox{ }
\left<
 \hat{{\cal U}}(t',t)^\dag
 \hat \mathrm{d}_{t'}^\dag
\hat{\cal H}(t') \hat{{\cal U}}(t',t)
\right. \nonumber \\ &  &  \left. \mbox{ }
+
  \hat{{\cal U}}(t',t)^\dag
\hat{\cal H}(t')  \hat \mathrm{d}_{t'}
\hat{{\cal U}}(t',t)
\right>_\mathrm{stoch}
\nonumber \\ & = &
\frac{-1}{T } \hat{\cal H}(t)
+
\frac{1}{T } \int_0^t \mathrm{d} t' \;
\left<
\hat{{\cal U}}(t',t)^\dag
( \partial_{t'}\hat{{\cal H}}(t'))   \hat{{\cal U}}(t',t)
\right>_\mathrm{stoch}
\nonumber \\ & \equiv &
\hat S_\mathrm{st}(t)
-
\int_0^t \mathrm{d} t' \;
\left<
\hat{{\cal U}}(t',t)^\dag
\hat{\dot S}\,\!^0_\mathrm{st}(t')
 \hat{{\cal U}}(t',t)
 \right>_\mathrm{stoch}
\nonumber \\ & \equiv &
\hat S_\mathrm{st}(t)
+ \hat S_\mathrm{dyn}(t).
\end{eqnarray}
%The first equality shows explicitly that this is an Hermitian operator.
%The second equality follows an integration by parts.
The steps in this derivation will be justified and interpreted below,
but first it is noted that
in the final equality the non-equilibrium reservoir entropy operator
has been split into  so-called static and dynamic parts.
This result holds in general
not just for the  mechanical non-equilibrium case
in which it was derived.

The static part is essentially the equilibrium expression
that one would write down for the system.
It gives predominantly the structure,
and it is insensitive to the direction of time
(i.e.\ the direction of the molecular velocities),
$ \hat S_\mathrm{st}(t)^* =  \hat S_\mathrm{st}(t)$.
For the  non-equilibrium mechanical system it is
just the Maxwell-Boltzmann form,
$ \hat S_\mathrm{st}(t) = - \hat{\cal H}(t)/T $.

The dynamic part part of the non-equilibrium reservoir entropy
is sensitive to the direction of time,
$ \hat S_\mathrm{dyn}(t)^* \ne  \hat S_\mathrm{dyn}(t)$.
Explicitly it is
\begin{equation} \label{Eq:^Sdyn}
\hat S_\mathrm{dyn}(t)
=
- \int_0^t \mathrm{d} t' \; \left<
\hat{{\cal U}}(t',t)^\dag
\hat{\dot S}\,\!^0_\mathrm{st}(t')
\hat{{\cal U}}(t',t)
\right>_\mathrm{stoch} .
\end{equation}
In essence this subtracts the total adiabatic change
in the static part of the reservoir entropy up to the present time.
The adiabatic derivative of the static part of the reservoir entropy is
\begin{equation}
\hat{\dot S}\,\!^0_\mathrm{st}(t)
=
\partial_t \hat S_\mathrm{st}(t)
- \frac{1}{i\hbar} \hat{\cal H}(t) \hat S_\mathrm{st}(t)
+ \frac{1}{i\hbar} \hat S_\mathrm{st}(t)  \hat{\cal H}(t).
\end{equation}
For the  non-equilibrium mechanical system this is just
$\hat{\dot S}\,\!^0_\mathrm{st}(t) = - \partial_t \hat{\cal H}(t)/T$.

Returning to the discussion of Eq.~(\ref{Eq:hatSr(t)}),
the passage from the first equality to the second equality
follows an integration by parts
and the fact that $ \hat{{\cal U}}(t,t) = \hat{\mathrm I}$.
Due to the conservation of energy on an adiabatic trajectory,
the partial time derivative of the energy operator
is equal to its adiabatic time derivative,
$ \partial_{t}\hat{{\cal H}}(t) = \hat{\dot {\cal H}}\,\!^0(t)$.
One can see therefore in the penultimate equality %of Eq.~(\ref{Eq:hatSr(t)})
that the integrand is the adiabatic rate of change
of the sub-system energy at time $t'$,
and so the difference between the first term on the left hand side
of the penultimate equality and the integral
is the change in energy of the sub-system induced
by the reservoir, all divided by temperature.
This is in accord with the physical interpretation offered
following Eq.~(\ref{S_r(psi[t])}) above.

In the penultimate equality of Eq.~(\ref{Eq:hatSr(t)}) has been defined
the static part of the reservoir entropy operator
for a mechanical non-equilibrium system,
\begin{equation}
\hat S_\mathrm{st}(t)
=
\frac{-1}{T } \hat{\cal H}(t) .
\end{equation}
This of course is just the instantaneous reservoir entropy operator
for a thermal equilibrium system.
The reason for casting the reservoir entropy operator
for a non-equilibrium system in the above terms of the equilibrium
entropy operator is that the formalism carries through
with only minor changes
for a thermodynamic non-equilibrium system.
A thermodynamic  non-equilibrium systems
consists of a sub-system sandwiched between two reservoirs
that apply a thermodynamic gradient.
This is the configuration that typically gives steady state flows;
heat flow, electric current, hydrodynamic flow,
and diffusive flux are common examples.
Explicitly, for steady heat flow
\begin{equation}
\hat S_\mathrm{st}(t)
=
\frac{-\hat E_0}{T_0 }  - \frac{\hat E_1}{T_1 } .
\end{equation}
where $T_0$ is essentially the average temperature of the two reservoirs
and $T_1$ is essentially the temperature gradient
between the two reservoirs,
and where $\hat E_0 \equiv \hat{\cal H}$ and $\hat E_1$
are the zeroth and first energy moment operators, respectively.
Whereas for a mechanical system with time-independent Hamiltonian
the adiabatic evolution of the energy operator
%(equivalently, the zeroth energy moment operator)
vanishes,
for steady heat flow %(and for other thermodynamic non-equilibrium systems)
the adiabatic evolution of the first energy moment
%(or other relevant first moment)
is non-zero.
See Refs~[\onlinecite{NETDSM, AttardI,AttardIII}] for full details
in the classical case.

The physical interpretation of the third equality of Eq.~(\ref{Eq:hatSr(t)})
is essentially the same as the interpretation of the second equality,
but it is a little more general because it encompasses both
mechanical and thermodynamic non-equilibrium systems.
The static reservoir entropy is expressed solely in terms of sub-system
properties, so it in itself is not the actual reservoir entropy.
However, subtracting the total adiabatic change of the static reservoir entropy
from the current value of the  static reservoir entropy
gives the change (from its value at $t=0$)
in the static reservoir entropy that is due solely to the interactions
with the reservoir.
It is this quantity that is the actual (change in the) reservoir entropy.

Of the two contributions to the reservoir entropy
of the non-equilibrium system,
the static part dominates for the structure,
but the dynamic term is essential to get
the irreversible aspects of a non-equilibrium system correct.
This dynamic term is the non-trivial part of the
entropy for a non-equilibrium system.
The Green-Kubo relations,\cite{Onsager31a,Green54,Kubo66}
which express the transport coefficients
as an integral of certain equilibrium time correlation functions,
come directly from this dynamic part.\cite{AttardV,NETDSM,Attard14}

%%%%%%%%%%%%%%%%%%%%%%%%%%%%%%%%%%%%%%%%%%%%%%%%
\subsubsection{Non-Equilibrium  Probability Operator}

Using the non-equilibrium reservoir entropy operator,
Eq.~(\ref{Eq:hatSr(t)}),
the probability operator
for a mechanical or thermodynamic non-equilibrium quantum system
is
\begin{eqnarray} \label{Eq:wp(t)}
\hat \wp(t)
& = &
\frac{1}{Z(t)}
e^{\hat S_\mathrm{r}(t)/k_\mathrm{B}}
 \\ \nonumber & = &
\frac{1}{Z(t)}
e^{ \left\{ \hat S_\mathrm{st}(t)
+ \hat S_\mathrm{dyn}(t) \right\} /k_\mathrm{B} }
  \\ \nonumber & = &
\frac{1}{Z(t)}
\exp \frac{1}{k_\mathrm{B}}
\left\{ \rule{0cm}{.5cm}% strut:
\hat{S}_\mathrm{st}(t)
\right.  \\ \nonumber &  & \left. \mbox{ }
-
\int_0^t \mathrm{d} t' \;
\left<
\hat{{\cal U}}(t',t)^\dag
\hat{\dot S}\,\!^0_\mathrm{st}(t')
 \hat{{\cal U}}(t',t)
 \right>_\mathrm{stoch}
\right\} .
\end{eqnarray}
Here $Z(t)$ is just the normalizing partition function.
This relies upon the result derived in paper I:\cite{QSM1}
due to wave function collapse,
the probability operator is the exponential
of the entropy operator divided by Boltzmann's constant.
(That result was derived for a canonical equilibrium system.)
One persuasive reason to believe
that the present result is the correct expression
for the non-equilibrium probability operator
is that keeping only the leading order term
reduces the present expression to the  equilibrium probability operator,
$\hat \wp_\mathrm{equil} = e^{\hat S_\mathrm{st}/k_\mathrm{B}}/Z$.

The non-equilibrium statistical average of a one-time operator is
\begin{eqnarray} \label{Eq:<O(t)>}
\langle \hat O(t) \rangle_{T,t}
& = &
\int \mathrm{d} \underline \psi \;
\frac{\langle \psi | \hat \wp(t) \hat O(t) | \psi\rangle
}{\langle\psi|\psi\rangle}
\nonumber \\ & = &
\mbox{Tr}^{(1)} \hat \wp(t) \hat O(t) .
\end{eqnarray}

In view of the time evolution of the operator given above,
Eq.~(\ref{Eq:O(t,t0)}),
this can also be written
\begin{eqnarray}
\langle \hat O(t_2) \rangle_{T,t}
& = &
\mbox{Tr}^{(1)}   \hat \wp(t_1) \hat O(t_2,t_1)
 \\ \nonumber & = &
\mbox{Tr}^{(1)} \left< \hat \wp(t_1)
\hat{{\cal U}}(t_2,t_1)^\dag \hat O(t_2)  \hat{{\cal U}}(t_2,t_1)
\right>_\mathrm{stoch}
 \\ \nonumber & = &
\mbox{Tr}^{(1)} \left<
\hat{{\cal U}}(t_2,t_1) \hat \wp(t_1)
\hat{{\cal U}}(t_2,t_1)^\dag \hat O(t_2)
\right>_\mathrm{stoch} .
\end{eqnarray}
The final equality follows from the cyclic properties of the trace.
Comparing this with Eq.~(\ref{Eq:<O(t)>}) evaluated at $t_2$,
one sees that one must have
\begin{equation}
\hat \wp(t_2)
=
 \left< \hat{{\cal U}}(t_2,t_1) \hat \wp(t_1)
\hat{{\cal U}}(t_2,t_1)^\dag \right>_\mathrm{stoch} .
\end{equation}
This is the time evolution of the probability operator,
and is in agreement with Eq.~(\ref{Eq:wp1-stat}).

\comment{ %%%%%%%%%%%%%%%%%%%%%%%%%%%%%
This expression for the probability operator evolution
is similar to, but slightly different from,
that for the operator evolution,   Eq.~(\ref{Eq:O(t,t0)}),
$\hat O(t_2,t_1) =
\langle
\hat{{\cal U}}(t_2,t_1)^\dag \hat O(t_2) \hat{{\cal U}}(t_2,t_1)
\rangle_\mathrm{stoch}$.
The operator evolutions of course mean different things:
$\hat \wp(t_2)$ is to be used in conjunction with $\psi(t_2)$,
whereas $\hat O(t_2,t_1)$ is to be used in conjunction with $\psi(t_1)$.
} % end comment %%%%%%%%%%%%%%%%%%%%%%%%%%%%%%%%%%%%%%%%%%

The expression for the non-equilibrium probability operator
and for the non-equilibrium reservoir entropy operator are formally exact.
However, to be useful, an explicit expression
for the  time propagator is required.
This is the subject of \S \ref{Sec:S2+SDSE}.
First, however,
a fluctuation formulation of the non-equilibrium reservoir entropy
is developed.

%%%%%%%%%%%%%%%%%%%%%%%%%%%%%%%%%%%%%%%%%%%%%%%%%%%%%%%%%%%%%%%%%%%%%%%%%%
\subsection{Fluctuation Form for the Reservoir Entropy}

%%%%%%%%%%%%%%%%%%%%%%%%%%%%%%%%%%%%%%%
\subsubsection{Ground State Projection}

Label complete sets of microstates by either
a single Roman letter such as $n$,
or else by a Greek and Roman pair of letters,
such as $\alpha h$, where $\alpha$ is the principle quantum number,
and $h$ labels the degeneracy.
A single Greek letter signifies a macrostate,
which is a quantum state of an incomplete operator.

Let $\zeta^\mathrm{S}_{\alpha h}(t)$ be an eigenfunction
of the reservoir entropy, such that
$ \hat S_\mathrm{r}(t) | \zeta^\mathrm{S}_{\alpha h}(t)\rangle
= S_\alpha(t) | \zeta^\mathrm{S}_{\alpha h}(t)\rangle $,
%Similarly for the static part
%$ \hat S_\mathrm{st}(t) | \zeta^\mathrm{S,st}_{\alpha h}(t)\rangle
%= S_\alpha^\mathrm{st}(t) | \zeta^\mathrm{S,st}_{\alpha h}(t)\rangle $.
with $\alpha = 0, 1, 2, \ldots$.
The eigenstates of maximum entropy correspond to $\alpha = 0$,
so that $S_\alpha(t) < S_0(t)$ for  $\alpha > 0$.
%Similarly $S_\alpha^\mathrm{st}(t) < S_0^\mathrm{st}(t)$ for  $\alpha > 0$.
The sub-space corresponding to $\alpha = 0$
contains the most likely wave states,
and it can also be called the entropy ground state.
The projection operator for this is
\begin{equation}
\hat{\cal P}_0(t) \equiv \sum_h
|\zeta^S_{0h}(t) \rangle \langle \zeta^S_{0h}(t) | .
\end{equation}
The projector for the orthogonal sub-space, the excited sub-space, is
$\hat{\cal P}_{\!\!\perp}(t) \equiv \hat{\mathrm I} - \hat{\cal P}_0(t)  $.
%One can append the superscript `st' to these
%for the static part of the entropy.

In general an operator can be decomposed
into its projections onto the two sub-spaces,
\begin{eqnarray}
\hat O & = &
[ \hat{\cal P}_0  + \hat{\cal P}_{\!\!\perp} ]
\hat O
[ \hat{\cal P}_0  + \hat{\cal P}_{\!\!\perp} ]
\nonumber \\ & = &
\hat{\cal P}_0  \hat O \hat{\cal P}_0
+
\hat{\cal P}_0  \hat O  \hat{\cal P}_{\!\!\perp}
+
\hat{\cal P}_{\!\!\perp} \hat O \hat{\cal P}_0
+
\hat{\cal P}_{\!\!\perp} \hat O \hat{\cal P}_{\!\!\perp}
\nonumber \\ & \equiv &
\hat O_{00} + \hat O_{0\perp} + \hat O_{\!\perp 0} + \hat O_{\!\perp\perp} .
\end{eqnarray}
These are of course time dependent.
The entropy operator is block diagonal,
\begin{eqnarray}
\hat S_\mathrm{r}(t)
& = &
\hat S_{\mathrm{r},00}(t)
+
\hat S_{\mathrm{r},\perp\perp}(t)
\nonumber \\ & = &
S_0(t) \hat{\cal P}_0(t)
+
\hat S_{\mathrm{r},\perp\perp}(t).
\end{eqnarray}

%%%%%%%%%%%%%%%%%%%%%%%%%%%%%%%%%%%%%%
\subsubsection{Most Likely Trajectory}

Now is introduced what will be called the most likely trajectory,
$\overline \psi(t)$,
although the nomenclature is not entirely satisfactory.
One of two things will be meant by this:
either this will be the ground state projection
of the current wave state,
$| \overline \psi(t) \rangle = \hat{\cal P}_0(t) | \psi \rangle$,
or else over a short time interval
it is the most likely evolution of
the ground state projection of the starting wave state,
$| \overline \psi(t_2) \rangle
= \hat{\overline{\cal U}}(t_2,t_1) \, \hat{\cal P}_0(t_1) | \psi_1 \rangle$.
Hence one has
\begin{equation}
\hat S_\mathrm{r}(t)  | \overline \psi(t) \rangle
=
S_0(t) | \overline \psi(t) \rangle.
\end{equation}
In the first case this is exact,
and in the second case it is a good approximation
for a short  time interval.

The current wave state $\psi$
can be considered in terms of the departure from the current value
of the most likely trajectory.
Hence the fluctuation at time $t$ can be defined as
%One can also write $ \overline \psi(t) \equiv \psi_0(t)  $.
\begin{equation}
| \phi  \rangle
\equiv
|  \psi - \overline \psi(t)  \rangle .
%\equiv [ \hat{\mathrm I} - \hat{\cal P}_0  ] | \psi(t)  \rangle
%\equiv \hat{\cal P}_{\!\!\perp}(t) | \psi(t)  \rangle.
\end{equation}
This is dominated by the excited states of the current wave state,
$ | \phi  \rangle \approx
\hat{\cal P}_{\!\!\perp}(t) | \psi  \rangle$
if $ |\overline \psi(t) \rangle \approx
\hat{\cal P}_0(t) |  \psi\rangle $.
This would be exact if the current value of the most likely
trajectory was taken as the ground state projection of the wave state.
In general one expects the departure from the ground state
to be small, and so this definition of a fluctuation
provides the basis for an expansion of the entropy and other
statistical mechanical properties.

The reason for being flexible with the definition
the most likely trajectory is that  the formalism
is most useful in two different cases.
The first case is for one-time thermodynamic properties
such as the entropy,
and this involves an expansion
about the ground state projection of the current
wave function.
The second case is for two-time thermodynamic properties
such as the second entropy, the transition probability,
and the time propagator.
These involve two fluctuation expansions:
one about the ground state projection of the first wave function
and the other about the most likely trajectory at the second time
emanating from the first  ground state projection.
It turns out that the  ground state projection of the most likely evolution
is not equal to the  most likely  evolution of the  ground state projection.

%%%%%%%%%%%%%%%%%%%%%%%%%%%%%%%%%%%%%%%%%%%%%%%%%%%%%%%%%%%%%%
\subsubsection{Fluctuation Operator and the Thermodynamic Force}

Now the second derivative of the entropy will be given
in order to write it in fluctuation form.
First however a specifically quantum issue
that does not arise in classical statistical mechanics
must be addressed.
As shown in Ref.~[\onlinecite{QSM3}],
one must distinguish between the entropy
and the expectation value of the entropy operator.
This effects the form of the fluctuation matrix and the thermodynamic force.

As mentioned briefly in \S \ref{Sec:Sr(psi,t)},
one has to distinguish between the actual entropy
and the expectation entropy.
The actual entropy, or simply entropy, is
\begin{equation}
S_\mathrm{r}(\psi,t)
\equiv
k_\mathrm{B} \ln
\frac{\langle \psi | e^{ \hat S_\mathrm{r}(t) /k_\mathrm{B} } | \psi \rangle
}{\langle \psi | \psi \rangle}  .
\end{equation}
The expectation value of the entropy operator
(expectation entropy, for short) is
\begin{equation}
S_\mathrm{r}^{<>}(\psi,t)
=
\frac{\langle \psi | \hat S_\mathrm{r}(t) | \psi \rangle
}{\langle \psi | \psi \rangle} .
\end{equation}
If the sub-system is in an entropy macrostate,
$| \psi^\mathrm{S}_\alpha \rangle
= \sum_h \psi_h  | \zeta^\mathrm{S}_{\alpha h}(t) \rangle$,
%with
%$ \hat S_\mathrm{r}(t) | \zeta^\mathrm{S}_{\alpha h}(t) \rangle
%= S_{\alpha}(t) | \zeta^\mathrm{S}_{\alpha h}(t) \rangle$,
these are equal,
\begin{equation}
S_\mathrm{r}^{<>}(\psi^\mathrm{S}_\alpha,t)
=
S_\mathrm{r}(\psi^\mathrm{S}_\alpha,t)
=  S_{\alpha}(t) ,
\end{equation}
since $ \hat S_\mathrm{r}(t) | \psi^\mathrm{S}_\alpha \rangle
= S_{\alpha}(t) | \psi^\mathrm{S}_\alpha \rangle $.
They are also equal when linearization is valid,
 such as at high temperatures.

Below, wave state transitions  will be analyzed
via the second entropy with the object
of deriving the stochastic, dissipative equation of motion
for the non-equilibrium system.
The thermodynamic driving force will arise from the reduction condition
on the second entropy in fluctuation form.
As shown in Ref.~[\onlinecite{QSM3}],
the reduction condition applies to the entropy,
not to the expectation entropy.
(Regrettably,  the fluctuation and force operators derived
in Ref.~[\onlinecite{QSM3}] are based on the expectation entropy,
which is a linearization of the results subsequently given in
Ref.~[\onlinecite{QSM2}] and that are given here.)

In view of this,
the thermodynamic force is
\begin{equation}
\frac{\partial S_\mathrm{r}(\psi,t)}{ \partial \langle \psi |  }
 =
k_\mathrm{B}  \left[
\frac{e^{ \hat S_\mathrm{r}(t) /k_\mathrm{B} }}
{\langle \psi | e^{ \hat S_\mathrm{r}(t) /k_\mathrm{B} } | \psi \rangle }
%\right. \nonumber \\ && \left. \mbox{ }
- \frac{1}{\langle \psi | \psi \rangle}
\right] | \psi \rangle .
%\nonumber \\ &=&
%\frac{ k_\mathrm{B}  }{ N(\psi)e^{ S_\mathrm{r}(\psi,t)/k_\mathrm{B}}  }
% \nonumber \\ &&  \mbox{ } \times
% \left[ e^{ \hat S_\mathrm{r}(t) /k_\mathrm{B} }
%% \right. \nonumber \\ && \left. \mbox{ }
%-  e^{ S_\mathrm{r}(\psi,t)/k_\mathrm{B}} \hat \mathrm{I}
%\right] | \psi \rangle ,
\end{equation}
Hence the  entropy force operator is defined as
\begin{eqnarray} \label{Eq:^Sr'}
\hat S_\mathrm{r}'(\psi,t)
&  \equiv &
k_\mathrm{B}  \left[
\frac{ e^{ \hat S_\mathrm{r}(t) /k_\mathrm{B} } }
{\langle \psi | e^{ \hat S_\mathrm{r}(t) /k_\mathrm{B} } | \psi \rangle }
-
\frac{1}{\langle \psi | \psi \rangle} \hat \mathrm{I}
\right]
\nonumber \\ & = &
\frac{k_\mathrm{B}}{N(\psi)} \left[
\frac{ e^{ \hat S_\mathrm{r}(t) /k_\mathrm{B} } }
{ e^{ S_\mathrm{r}(\psi,t)/k_\mathrm{B} }  }
-  \hat \mathrm{I} \right] .
\end{eqnarray}
Because this depends upon the wave state
it is a non-linear operator.
%since
% $ S_\mathrm{r}(\psi,t)
%= k_\mathrm{B} \ln [
%\langle \psi | e^{ \hat S_\mathrm{r}(t) /k_\mathrm{B} } | \psi \rangle
%/\langle \psi |  \psi \rangle]$.

The second derivative gives
the entropy fluctuation operator,
\begin{eqnarray} \label{Eq:^Sr''}
\hat S_\mathrm{r}''(\psi,t) & \equiv &
\frac{\partial S_\mathrm{r}(\psi,t)
}{ \partial | \psi \rangle  \partial \langle \psi |}
 \\ & = &
k_\mathrm{B}  \left[
\frac{e^{ \hat S_\mathrm{r}(t) /k_\mathrm{B} }}
{\langle \psi | e^{ \hat S_\mathrm{r}(t) /k_\mathrm{B} } | \psi \rangle }
%\right. \nonumber \\ && \left. \mbox{ }
-
\frac{1}{\langle \psi | \psi \rangle} \hat \mathrm{I}
\right]
\nonumber \\ && \mbox{ }
- k_\mathrm{B}
\frac{e^{ \hat S_\mathrm{r}(t) /k_\mathrm{B} }| \psi \rangle
\langle \psi | e^{ \hat S_\mathrm{r}(t) /k_\mathrm{B} }  }
{\langle \psi | e^{ \hat S_\mathrm{r}(t) /k_\mathrm{B} } | \psi \rangle^2 }
%\nonumber \\ &&  \mbox{ }
+ k_\mathrm{B}
\frac{| \psi \rangle \langle \psi | }{\langle \psi | \psi \rangle^2 } .
\nonumber
\end{eqnarray}
This is also a non-linear operator.

These are to be evaluated on the most likely trajectory,
$\overline \psi(t)$,
which lies in the ground state sub-space,
$ \hat S_\mathrm{r}(t)  | \overline \psi(t) \rangle
= S_0(t) | \overline \psi(t) \rangle$.
In the ground state,
the final two terms cancel
and the fluctuation operator becomes
\begin{eqnarray} \label{Eq:^ol-Sr''}
\hat {\overline S}_\mathrm{r}\!''(t) & \equiv &
\hat S_\mathrm{r}''(\overline \psi(t),t)
\nonumber \\ & = &
\frac{ k_\mathrm{B} }{N(\overline \psi(t))} \left[
\frac{ e^{ \hat S_\mathrm{r}(t) /k_\mathrm{B} } }
{ e^{ S_0(t) /k_\mathrm{B} } }
-  \hat \mathrm{I} \right] .
\end{eqnarray}
This is equal to the entropy force operator
also evaluated in the entropy ground state,
$ \hat {\overline S}_\mathrm{r}\!''(t)
= \hat {\overline S}_\mathrm{r}\!'(t)
\equiv \hat S_\mathrm{r}'(\overline \psi(t),t)$.
By design,
the thermodynamic force vanishes on the most likely trajectory,
$\hat {\overline S}_\mathrm{r}\!''(t) | \overline \psi(t) \rangle
= \hat {\overline S}_\mathrm{r}\!'(t) | \overline \psi(t) \rangle
= | 0 \rangle$,
since this has been taken to lie in the ground state sub-space.
%(In fact, the force vanishes for any pure entropy macrostate
%in which the fluctuation operator is evaluated.)

Since the most likely trajectory lies in the ground state sub-space,
$S_\mathrm{r}(\overline \psi(t),t)$ is a maximum.
Hence one can expand the entropy to quadratic order
about the most likely trajectory, which is just fluctuation theory.
Recalling the definition of a fluctuation
as the departure from the most likely trajectory,
$ | \phi(t)  \rangle \equiv
|  \psi - \overline \psi(t)  \rangle $,
the (first) entropy in fluctuation form is
\begin{equation} \label{Eq:S1-fluct}
S_\mathrm{r}(\psi,t)
=
\overline S_\mathrm{r}(t)
+
\langle \phi | \hat {\overline S}_\mathrm{r}\!''(t) | \phi \rangle
+ {\cal O}(\phi^3).
\end{equation}
It is most useful in this case to take
$|  \overline \psi(t)  \rangle = \hat{\cal P}_0(t) |  \psi  \rangle$.
The linear terms,
which correspond to the thermodynamic force,
vanish because this is an expansion about the ground state.

Differentiating this expression
and comparing it to the derivative above,
one can write  the thermodynamic force for the wave state $\psi$
in several equivalent ways,
\begin{equation} \label{Eq:Sr'psi=Sr''phi}
\frac{\partial S_\mathrm{r}(\psi,t)
}{ \partial \langle \psi |}
=
\hat {\overline S}_\mathrm{r}\!''(t) \,  | \phi \rangle
=
\hat {\overline S}_\mathrm{r}\!''(t) \,  | \psi \rangle
=
\hat {\overline S}_\mathrm{r}\!'(t) \,  | \psi \rangle .
%\nonumber \\ & \approx &
%%\hat{\cal P}_\perp(t)
%\hat S_\mathrm{r}'(\psi,t) \,
%%\hat{\cal P}_\perp(t)
% | \psi \rangle.
\end{equation}
These neglect terms ${\cal O}(\phi^2)$.
%The first three equalities are exact.
%The fourth equality is an approximation that neglects terms ${\cal O}(\phi^2)$,
%and that does not vanish  in the ground state.
%There is probably not much point in it.
%Also
%$\hat {\overline S}_\mathrm{r}\!''(t) \,  | \phi \rangle
%= \hat {\overline S}_\mathrm{r}\!''(t) \,  | \psi \rangle $,
%but this is not required.

%One can define similar force and fluctuation operators
%for the expectation entropy.
%These are not required here,
%but in the future they could be explored
%as  a possible way to simplify  the computational problem.

%%%%%%%%%%%%%%%%%%%%%%%%%%%%%%%%%%%%%%%%%%%%%%%%%%%%%%%%%%%%%%
\subsubsection{Static Entropy Force Operator} \label{Sec:Sst}

In the classical version of the non-equilibrium theory,
it was found fruitful,
both conceptually and computationally,
to replace the full entropy fluctuation matrix
by the static part of the entropy fluctuation matrix.\cite{NETDSM}
A similar replacement will be explored here,
although it is not required until Eq.~(\ref{Eq:olpsi2}) below.

It will be recalled that the entropy operator
is the sum of a static part and a dynamic part,
\begin{equation}
\hat S_\mathrm{r}(t) =
\hat S_\mathrm{st}(t) + \hat S_\mathrm{dyn}(t) .
\end{equation}
Since the expectation entropy is a linear function
of the entropy operator,
it can also be written as the sum of  static  and  dynamic parts.
However, as mentioned above,
it is the entropy $ S_\mathrm{r}(\psi,t)$ rather than the entropy
expectation value $ S_\mathrm{r}^{<>}(\psi,t)$
that features in the reduction condition
and that will appear in the stochastic, dissipative equation of motion.
The entropy is a non-linear function of the entropy operator,
and it is not possible to split it into pure static and dynamic parts.
One can however \emph{define} the static part of the
entropy to be
\begin{equation}
S_\mathrm{st}(\psi,t)
\equiv S_\mathrm{r}(\psi,t,[\hat S_\mathrm{dyn}\equiv 0]),
\end{equation}
and the dynamic part to be the remainder
\begin{equation}
S_\mathrm{dyn}(\psi,t)
\equiv S_\mathrm{r}(\psi,t) - S_\mathrm{st}(\psi,t) .
\end{equation}
Note that the dynamic part contains static contributions,
$S_\mathrm{dyn}(\psi,t)
\ne S_\mathrm{r}(\psi,t,[\hat S_\mathrm{st}\equiv 0])$.

As mentioned above,
in the classical case it was found useful
to replace the full entropy fluctuation matrix
by the static part of the entropy fluctuation matrix.\cite{NETDSM}
In the present case, the analogous replacement is
\begin{equation} \label{Eq:Sr''-Sst''}
\hat {\overline S}_\mathrm{r}\!\!''(t)
\approx \hat {\overline S}_\mathrm{st}\!\!\!''(t) ,
\end{equation}
where the static part of the entropy fluctuation operator
is defined to be
\begin{equation}
\hat {\overline S}_\mathrm{st}\!\!\!''(t) \equiv
\hat {\overline S}_\mathrm{st}\!\!'(t) \equiv
\frac{ k_\mathrm{B} }{N(\overline \psi(t))} \left[
\frac{ e^{ \hat S_\mathrm{st}(t) /k_\mathrm{B} } }
{ e^{ S_0(t) /k_\mathrm{B} } }
-  \hat \mathrm{I} \right] .
\end{equation}
Note that here and below the over line signifies quantities
evaluated in the ground state of the entropy,
not the ground state of the static part of the entropy.
Specifically, $ \hat S_\mathrm{r}(t) | \overline \psi(t)\rangle
= S_0(t)  | \overline \psi(t)\rangle $.
This is the reason that  it is $S_0(t)$, not $S_0^\mathrm{st}(t)$,
that appears in this definition of $\hat {\overline S}_\mathrm{st}\!\!'(t)$.
Note that this latter operator
is not guaranteed to be negative definite,
but  one nevertheless expects it to be at least approximately so.

The preceding equation gives the first and second derivatives
of the static part of the entropy
evaluated in the reservoir entropy ground state.
This is precisely what is required for the fluctuation expansion
of the static part of the entropy about the entropy ground state,
\begin{eqnarray}
S_\mathrm{st}(\psi,t) & = &
{\overline S}_\mathrm{st}(t)
+
\langle \overline \psi(t) |
\hat {\overline S}_\mathrm{st}\!\!'(t) | \phi \rangle
+
\langle \phi | \hat {\overline S}_\mathrm{st}\!\!'(t)
|  \overline \psi(t) \rangle
\nonumber \\ && \mbox{ }
+
\langle \phi | \hat {\overline S}_\mathrm{st}\!\!\!''(t) | \phi \rangle
+ {\cal O}(\phi^3) .
\end{eqnarray}
The linear terms do not vanish
because the point of expansion, $\overline \psi(t)$,
is not a maximum  of the static entropy operator.
Combining this with the fluctuation expansion of the entropy,
Eq.~(\ref{Eq:S1-fluct}),
then by definition the dynamic part of the reservoir entropy
has expansion
\begin{eqnarray}
S_\mathrm{dyn}(\psi,t)
& \equiv &
S_\mathrm{r}(\psi,t) -
S_\mathrm{st}(\psi,t)
\nonumber \\& = &
\overline S_\mathrm{r}(t) - {\overline S}_\mathrm{st}(t)
-
\langle \overline \psi(t) |
\hat {\overline S}_\mathrm{st}\!\!'(t) | \phi \rangle
\nonumber \\ && \mbox{ }
-
\langle \phi | \hat {\overline S}_\mathrm{st}\!\!'(t)
|  \overline \psi(t) \rangle
+ {\cal O}(\phi^3) .
\end{eqnarray}
The quadratic terms cancel
upon using the above replacement,
$ \hat {\overline S}_\mathrm{r}\!''(t)
\approx
\hat {\overline S}_\mathrm{st}\!\!''(t)$.
Hence the dynamic part of the thermodynamic force
is to leading order constant in the excited sub-space of wave space,
\begin{eqnarray} \label{Eq:Sdyn'}
\frac{\partial S_\mathrm{dyn}(\psi,t)
}{ \partial \langle \psi |}
& = &
- \hat {\overline S}_\mathrm{st}\!\!'(t) |  \overline \psi(t) \rangle
+ {\cal O}(\phi^2) .
\end{eqnarray}
This is entirely analogous to the classical result
(see Eq.~(8.26) of Ref.~[\onlinecite{NETDSM}],
or  Eq.~(20) of Ref.~[\onlinecite{Attard14}]).
This is only constant as far as the excited sub-space is concerned,
since $|  \overline \psi(t) \rangle
= \hat{\cal P}_0(t) \, |  \psi \rangle$.

Differentiating these two fluctuation forms,
the consequent thermodynamic force is
\begin{eqnarray} \label{Eq:Sr'-Sst'}
%\lefteqn{
\frac{\partial S_\mathrm{r}(\psi,t)
}{ \partial \langle \psi |}
%} \nonumber \\
& = &
\hat {\overline S}_\mathrm{st}\!\!'(t) |  \overline \psi(t) \rangle
+
 \hat {\overline S}_\mathrm{st}\!\!\!''(t) | \phi \rangle
- \hat {\overline S}_\mathrm{st}\!\!'(t) |  \overline \psi(t) \rangle
%+ {\cal O}(\phi^2)
\nonumber \\ & = &
 \hat {\overline S}_\mathrm{st}\!\!\!''(t) | \phi \rangle
\nonumber \\ & = &
\hat {\overline S}_\mathrm{st}\!\!'(t) \hat{\cal P}_\perp(t) | \psi \rangle.
%+ {\cal O}(\phi^2),
\end{eqnarray}
The neglected terms here are of quadratic order in the fluctuation.
The penultimate equality could have been written down directly by
making the static fluctuation operator replacement,
Eq.~(\ref{Eq:Sr''-Sst''}), in the original  fluctuation expansion of
the entropy, Eq.~(\ref{Eq:S1-fluct}), and differentiating.
Part of the point of the exercise, however, was to explore the implications
for the dynamic part of the entropy, and also to show the consonance
with the classical results.
Additionally, it gives certainty about the exact form
for the static fluctuation operator, which is always desirable.
The final equality uses the fact that
$ \hat {\overline S}_\mathrm{st}\!\!\!''(t) =
  \hat {\overline S}_\mathrm{st}\!\!'(t)$.
It also assumes that $| \overline \psi(t)\rangle
= \hat{\cal P}_0(t)  | \psi \rangle$.
This form will prove
useful in Eq.~(\ref{Eq:olpsi2}) \emph{et seq.}\ below.

Using the static part of the fluctuation operator
in place of the full fluctuation operator, Eq.~(\ref{Eq:Sr''-Sst''}),
might appear to be an approximation,
but several justifications can be offered.
In the first place, fluctuations about
the non-equilibrium state are determined by the current molecular structure,
and that these fluctuations
have the same symmetries as equilibrium fluctuations.
There is an abundance of computer simulation
data for classical system that show that the  fluctuations
in a non-equilibrium system are identical to those
in the corresponding local equilibrium system.
\cite{AttardV,AttardVIII,AttardIX,Attard09}
%This approximation has been found to be accurate
%in computer simulation tests for both mechanical
%\cite{AttardVIII,AttardIX,Attard09}
%and thermodynamic
%\cite{AttardV,AttardIX} non-equilibrium systems.
Further, this replacement appears necessary on physical grounds,
namely that  in the dissipative Schr\"odinger equation,
this term comes from the reservoir--sub-system interactions
and it is the static part of the entropy operator
that fully reflects such interactions
(c.f.\ the discussion in the conclusion of Ref.~[\onlinecite{Attard14}]).
Finally, the point entropy derived above is based upon
the average trajectory
(more precisely, the average of the square of the stochastic propagator),
and its gradient gives the difference in this average value.
What is required for the dissipative Schr\"odinger equation,
is the change in reservoir entropy on an actual trajectory,
and this is given exactly by the gradient in the static part of the
entropy.
This matter is further discussed at the end of \S \ref{Sec:DSE} below.

%\newpage %$\;$ \newpage
%%%%%%%%%%%%%%%%%%%%%%%%%%%%%%%%%%%%%%%%%%%%%%%%%%%%%%%%%%%%%%%%%%%%%%%%%%%%%%%
%                                                                             %
                \section{Transitions and Motion in Wave Space}
\setcounter{equation}{0} \label{Sec:S2+SDSE}
%                                                                             %
%%%%%%%%%%%%%%%%%%%%%%%%%%%%%%%%%%%%%%%%%%%%%%%%%%%%%%%%%%%%%%%%%%%%%%%%%%%%%%%

\subsection{Second Entropy and the Most Likely Transition}

The second entropy is the entropy for transitions.
Accordingly, maximizing it determines
the most likely trajectory and gives the most likely propagator.
In \S 3A of Paper II\cite{QSM2}
and in \S 3 of Paper III\cite{QSM3}
of the present series,
the second entropy for the quantum equilibrium case was analyzed.
The present non-equilibrium case utilizes the same analysis and notation,
the main difference being that microscopic reversibility
does not here hold.
The approach that follows is essentially the quantum version
of the non-equilibrium classical analysis given
in \S 8.3.2 of Ref.~[\onlinecite{NETDSM}]
or \S\S IIB and III of Ref.~[\onlinecite{Attard14}].

The focus is on the transition $\{\psi_1,t_1\} \rightarrow \{\psi_2,t_2\}$.
If $t_2>t_1$, then this is a physical transition,
(i.e.\ it answers the question  ---where will the system go to?),
and if $t_2<t_1$, then this is a mathematical transition,
(i.e.\ it answers the question ---where did the system come from?).

Recall that the fluctuation of the sub-system wave function is
$\phi(t) \equiv \psi(t) - \overline \psi(t)$.
In the above, the most likely trajectory was taken to be the ground state
projection of the current wave function,
$|\overline \psi(t) \rangle = \hat{\cal P}_0(t) | \psi \rangle$.
However the above dealt with one-time quantities,
and here the concern is with two-time quantities,
so some care is required.
Here the most likely trajectory at the initial time is
taken to be the ground state projection of the initial wave function,
$|\overline \psi(t_1) \rangle = \hat{\cal P}_0(t_1) | \psi_1 \rangle$,
and at the final time it is taken to be the most likely evolution
of this,
$|\overline \psi(t_2) \rangle =
\hat{\overline{\cal U}}t_2,t_1) \,|\overline \psi(t_1) \rangle $.
(The outcome of this section will be the explicit form
for the time propagator.)
Hence the fluctuations are
\begin{eqnarray} \label{Eq:phi1,2}
| \phi_1 \rangle
& \equiv &
%\hat{\cal P}_\perp(t_1) | \psi_1 \rangle ,
| \psi_1 \rangle - \hat{\cal P}_0(t_1) | \psi_1 \rangle ,
\nonumber \\
\mbox{ and } | \phi_2 \rangle
& \equiv &
| \psi_2 %- \overline \psi(t_2|\overline \psi_1,t_1)
\rangle
- \hat{\overline{\cal U}}t_2,t_1) \hat{\cal P}_0(t_1) | \psi_1 \rangle .
\end{eqnarray}
It is not necessary to formulate these in a more symmetric fashion,
as the symmetry of the quadratic form given next  will shortly be broken.
It is to be noted that $\phi_1$ lies entirely in the excited state,
and, since $|t_{21}| \rightarrow 0$,
it is also true that  $\overline \psi(t_2)$
is either fully or predominantly in the ground state.
It is actually $\overline \psi(t_1)$ rather than $\overline \psi(t_2)$
that must lie in the ground state because
the non-linear thermodynamic force operator
that is required below is evaluated at $t_1$ and at
$|\overline \psi(t_1) \rangle = \hat{\cal P}_0(t_1) | \psi_1 \rangle$.

Assume that the second entropy for the transition
has the quadratic fluctuation form
\begin{eqnarray} \label{Eq:S2-flucn}
\lefteqn{
S^{(2)}( \psi_2, t_2; \psi_1, t_1)
} \nonumber \\
& = &
%\left<  \phi_2 ,  \phi_1 \right|
%\hat {\cal A}^{(2)}(t_{21},t)
%\left| \phi_1 ,  \phi_2 \right>
%+ \frac{1}{2}
%\left[ \overline S_\mathrm{r}(t_2) + \overline S_\mathrm{r}(t_1) \right]
%\nonumber \\ \nonumber & = &
\left< \phi_2 \right|  \hat a(t_{21},t) \left| \phi_2 \right>
%\nonumber \\ &&  \mbox{ }
+
\left< \phi_1 \right|  \hat c(t_{21},t) \left| \phi_1 \right>
\nonumber \\  &&  \mbox{ }
+
\left<  \phi_2 \right|  \hat b(t_{21},t) \left|  \phi_1 \right>
+
\left<  \phi_1 \right|  \hat b(t_{21},t)^\dag \left|  \phi_2 \right>
\nonumber \\ && \mbox{ }
+ \frac{1}{2}
\left[ \overline S_\mathrm{r}(t_2) + \overline S_\mathrm{r}(t_1) \right] .
\end{eqnarray}
Here $t \equiv [t_1+t_2]/2$ is the midpoint,
and $t_{21} \equiv t_2 - t_1$
is the first time in the argument minus the second.
Usually the time argument $t$ will be suppressed.
The first entropies that appear here
are $\overline S_\mathrm{r}(t_i)
\equiv S_\mathrm{r}(\overline \psi(t_i), t_i)$, $i=1,2$,
which are equal either exactly or  approximately
to the ground state entropy $ S_0(t_i)$.

%The negative definite operator $\hat {\cal A}^{(2)}(t_{21},t)$
%(equivalently its component operators
%$\hat a(t_{21},t) $, $\hat b(t_{21},t) $, and $\hat c(t_{21},t) $)
%is the second entropy fluctuation operator,
%and, after its properties have been established,
%it will be used to give the second entropy operator itself,
%and hence the transition probability operator.

The origin of the final time-dependent constant
is derived in \S 8.3.1 of Ref.~[\onlinecite{NETDSM}]
and in \S 4 of Ref.~[\onlinecite{QSM3}].
(The analogous constant was neglected in transforming the trajectory
entropy operator to the point entropy operator,
Eq.~(\ref{Eq:^Sr(t';t)=>^Sr(t)}),
as was discussed in the final paragraph of  \S \ref{Sec:Red-Traj-S}.)
It is related to the reduction condition,\cite{NETDSM,AttardII}
%---equivalent to the condition that the transition probability
%must reduce to the probability of the initial state
%upon summing over all the final states---
which for the conditionally most likely state
$\overline \psi_2 \equiv \overline \psi(t_2|\psi_1,t_1)$
is
\begin{equation} \label{Eq:S2-red}
S^{(2)}( \overline \psi_2, t_2; \psi_1, t_1)
=
S_\mathrm{r}(\psi_1, t_1)
+ \left[ \overline S_\mathrm{r}(t_2) - \overline S_\mathrm{r}(t_1) \right]/2 .
\end{equation}
This corresponds to  Eq.~(8.38) of Ref.~[\onlinecite{NETDSM}]
and to Eq.~(4.4) of Ref.~[\onlinecite{QSM3}].
As shown in Ref.~[\onlinecite{QSM3}],
the reduction condition applies to the entropy,
not the expectation entropy.

The second entropy must reflect statistical symmetry,
$S^{(2)}( \psi_2, t_2; \psi_1, t_1) = S^{(2)}( \psi_1, t_1;\psi_2, t_2)$.
Hence
\begin{equation}
\hat a(t_{12}) = \hat c(t_{21})
, \mbox{ and }
\hat b(t_{12}) = \hat b(t_{21})^\dag .
\end{equation}
The second entropy must be real,
$S^{(2)}( \psi_2, t_2; \psi_1, t_1)^* = S^{(2)}( \psi_2, t_2; \psi_1, t_1)$.
Hence
\begin{equation}
\hat a(t_{21}) = \hat a(t_{21})^\dag
, \mbox{ and }
\hat c(t_{21}) = \hat c(t_{21})^\dag ,
\end{equation}
which is to say that they are self-adjoint.
The cross-term is real by design.

These symmetries also held in the equilibrium quantum case
treated in  \S 3A of Paper II.\cite{QSM2}
The only symmetry that does not hold in the present
non-equilibrium case is microscopic reversibility.
Hence much of the analysis of  \S 3A\cite{QSM2}
holds also here.

In detail,
since $ \overline \psi_2 \rightarrow \psi_1$ as $|t_{21}| \rightarrow 0$,
the second entropy must contain essentially a $\delta$-function singularity.
Hence the small $|t_{21}| $ expansions must be of the form
\begin{equation}
\hat a(\tau) =
\frac{1}{|t_{21}|} \hat a_{-1} + \frac{1}{t_{21}} \hat a_{-1}'
+
\hat a_{0} + \widehat \tau \hat a_{0}' + {\cal O}(t_{21}) ,
\end{equation}
\begin{equation}
\hat b(t_{21}) =
\frac{1}{|t_{21}|} \hat b_{-1} + \frac{1}{t_{21}} \hat b_{-1}'
+
\hat b_{0} + \widehat \tau \hat b_{0}' + {\cal O}(t_{21}) ,
\end{equation}
and
\begin{equation}
\hat c(t_{21}) =
\frac{1}{|t_{21}|} \hat c_{-1} + \frac{1}{t_{21}} \hat c_{-1}'
+
\hat c_{0} + \widehat \tau \hat c_{0}' + {\cal O}(t_{21}) ,
\end{equation}
with $\widehat \tau \equiv \mbox{sign }t_{21} = t_{21}/|t_{21}|$.

The  reason why the non-analytic terms appear
(i.e.\ those containing $|t_{21}|$ and $\widehat \tau$)
is that these are necessary to yield the irreversible behavior
that is characteristic of all thermodynamic evolution.
One concludes that this is not a Taylor expansion
for an infinitesimal time step,
since this would only ever yield analytic terms,
but rather an expansion for small but finite time steps
that is a re-summation of an infinite order  Taylor expansion.
The validity of beginning the expansion
with terms ${\cal O} (t_{21}^{-1})$ can be judged by the consequences;
amongst other things it yields a physically plausible
stochastic Schr\"odinger equation with a conventional velocity
for the wave function.

From the symmetries given above,
$\hat a(t_{21}) = \hat a(t_{21})^\dag  $,
and $ \hat a(t_{21}) = \hat c(-t_{21})$,
one can see that
the unprimed $\hat a$ are self-adjoint
and equal the unprimed $\hat c$,
and the primed $\hat a$ are self-adjoint
and equal the negative of the primed $\hat c$.
Also,
since $\hat b(t_{21}) = \hat b(-t_{21})^{\dag}  $,
the unprimed $\hat b$ are self-adjoint,
and the primed $\hat b$ are anti-self-adjoint.
(In quantum mechanics, the words Hermitian operator
and self-adjoint operator are used synonymously.)\cite{Merzbacher70}

Since $ \overline \psi_2 \rightarrow \psi_1$ as $|t_{21}| \rightarrow 0$,
to leading order $ \hat a(t_{21}) = - \hat b(t_{21})$,
which implies that
\begin{equation}
\hat a_{-1} = - \hat b_{-1}  \equiv -\hat \lambda^{-1}
, \mbox{ and }
\hat a_{-1}' = -\hat b_{-1}' = \hat 0 .
\end{equation}
From the symmetry relations,
$\hat \lambda$ is an Hermitian operator that is positive definite
(because the second entropy must be negative definite).
For reasons that will become clear shortly,
this will be called the dissipative operator,
although it could equally well be called the drag operator,
or the friction operator.
The primed coefficients individually  vanish because
$\hat a_{-1}'$ is self-adjoint
and $\hat b_{-1}'$ is anti-self-adjoint.
With these, the small time expansions  read
\begin{equation}
\hat a(t_{21}) =
\frac{ - 1}{|t_{21}|} \hat \lambda^{-1}
+ \hat a_{0} + \widehat \tau \hat a_{0}' + {\cal O}(t_{21}) ,
\end{equation}
\begin{equation}
\hat b(t_{21}) =
\frac{1}{|t_{21}|} \hat \lambda^{-1}
+ \hat b_{0} + \widehat \tau \hat b_{0}' + {\cal O}(t_{21}),
\end{equation}
and
\begin{equation}
\hat c(t_{21}) =
\frac{-1}{|t_{21}|} \hat \lambda^{-1}
+ \hat a_{0} - \widehat \tau \hat a_{0}' + {\cal O}(t_{21}) .
\end{equation}
Here $\hat \lambda$, $\hat a_{0}$,  $\hat a_{0}'$, and $\hat b_{0}$
are Hermitian, and $\hat b_{0}'$ is anti-Hermitian.
These all depend upon $t$ (not shown).

Maximizing the second entropy by setting
its derivative with respect to $\langle \phi_2 |$ to zero,
one obtains the conditional most likely state as
\begin{eqnarray} \label{Eq:olphi2ab}
| \overline \phi_2 \rangle
& = &
- \hat a(t_{21})^{-1} \hat b(t_{21}) | \phi_1 \rangle
\nonumber \\ & = &
|  \phi_1 \rangle
+ t_{21} \hat \lambda \left[ \hat a_0' + \hat b_0' \right]
|  \phi_1 \rangle
+ | t_{21} | \hat \lambda \left[ \hat a_0 + \hat b_0 \right]
|  \phi_1 \rangle
\nonumber \\ & & \mbox{ }
+ {\cal O}(t_{21}^2) .
\end{eqnarray}
The left hand side is more precisely written
$| \overline \phi(t_{21}|\phi_1,t) \rangle$.

From the reduction condition Eq.~(\ref{Eq:S2-red})
and the fluctuation expression Eq.~(\ref{Eq:S1-fluct})
one has
\begin{eqnarray}
S^{(2)}( \overline \psi_2, t_2; \psi_1, t_1)
& = &
S_\mathrm{r}(\psi_1, t_1)
+ \frac{1}{2}
\left[ \overline S_\mathrm{r}(t_2) - \overline S_\mathrm{r}(t_1) \right]
\nonumber \\ & = &
\langle \phi_1 | \hat {\overline S}_\mathrm{r}\!\,''(t_1) | \phi_1  \rangle
\nonumber \\ & & \mbox{ }
+ \frac{1}{2}
\left[ \overline S_\mathrm{r}(t_2) + \overline S_\mathrm{r}(t_1) \right] .
\end{eqnarray}
Evaluating the left hand side
using the preceding expression for $| \overline \phi_2 \rangle$
and its conjugate $ \langle\overline \phi_2  |$
in the fluctuation expression for  the second entropy,
Eq.~(\ref{Eq:S2-flucn}),
one sees that the constant term  in the final equality here
equals the constant term in the second entropy, Eq.~(\ref{Eq:S2-flucn}).
What remains on the left hand side is the expectation value
of an operator
(that is a function of $\hat a(t_{21})$,  $\hat b(t_{21})$,
and   $\hat c(t_{21})$)
in the wave state $\phi_1$,
which when set equal to the same expectation value on the final equality here
yields the explicit reduction condition
\begin{equation}
\hat c(t_{21},t)
- \hat b(t_{21},t)^\dag \hat a(t_{21},t)^{-1} \hat b(t_{21},t)
=
\hat {\overline S}_\mathrm{r}\!\,''(t_1) .
\end{equation}
The left hand side is clearly Hermitian, as it must be.
This equality has to hold for all values of $t_{2}$.
Since this is formally identical to the corresponding equilibrium result,
the small time expansion given in Eq.~(3.21) of Ref.~[\onlinecite{QSM2}]
must hold, and so the expansion coefficients of the operators
must reduce to
\begin{equation}
\hat a_0(t) + \hat b_0(t)
=
 \frac{1}{2} \hat {\overline S}_\mathrm{r}\!\,''(t)
=
 \frac{1}{2} \hat {\overline S}_\mathrm{r}\!\,'(t) .
\end{equation}
Recall from Eq.~(\ref{Eq:^ol-Sr''}) that in the ground state,
the entropy fluctuation operator equals the entropy force operator,
$ \hat {\overline S}_\mathrm{r}\!''(t)
= \hat {\overline S}_\mathrm{r}\!'(t)
\equiv \hat S_\mathrm{r}'(\overline \psi(t),t)$.
%To leading order, it makes no difference
%whether this  is evaluated at $t$, $t_1$, or $t_2$.

With this result, the irreversible operator (i.e.\ the operator
proportional to $|t_{21}|$) for the  most likely evolution of the
fluctuation, Eq.~(\ref{Eq:olphi2ab}), is
\begin{eqnarray}
%\hat{\overline R}_\phi(t_{21},t) & \equiv &
| t_{21} | \hat \lambda(t)
\left[ \hat a_0(t) + \hat b_0(t) \right]
%\nonumber \\
& = &
\frac{| t_{21}| }{2}  \hat \lambda(t) \hat {\overline S}_\mathrm{r}\!\,'(t) .
\end{eqnarray}
When this is inserted into the evolution equation Eq.~(\ref{Eq:olphi2ab}),
one sees that the irreversible term includes the factor
$ \hat {\overline S}_\mathrm{r}\!\,'(t) | \phi_1 \rangle
= \hat {\overline S}_\mathrm{r}\!\,'(t) | \psi_1 \rangle$.
This is the thermodynamic force or entropy gradient,
Eq.~(\ref{Eq:Sr'psi=Sr''phi}).
This term drives the current wave state of the sub-system, $\psi_1$,
toward the ground state.
This term ultimately gives the change in entropy during the transition,
which is the origin of the name `dissipative operator'
for $\hat \lambda(t)$.

As in the equilibrium case treated in Paper II
(see Eq.~(3.32) of Ref.~[\onlinecite{QSM2}])
the reversible term in  Eq.~(\ref{Eq:olphi2ab})
(the one proportional to $t_{21}$)
must contain the adiabatic evolution,
\begin{equation} \label{Eq:a0'+b0'=^H}
t_{21}\hat \lambda \left[ \hat a_0' + \hat b_0' \right]
=
\frac{t_{21}}{i\hbar}\hat{\cal H}(t) .
\end{equation}
Any reversible reservoir contribution is expected to be negligible
compared to this.
For completeness, the Hermitian conjugate of this is
\begin{equation}
\left[ \hat a_0' - \hat b_0' \right] \hat \lambda
=
\frac{-1}{i\hbar}\hat{\cal H}(t) .
\end{equation}

%\newpage
%%%%%%%%%%%%%%%%%%%%%%%%%%%%%%%%%%%%%%%%%%%%%%%%%%%%%
\subsection{Dissipative Schr\"odinger Equation}

With these results the conditional most likely fluctuation is
\begin{eqnarray} \label{Eq:olphi2}
| \overline \phi_2 \rangle
& = &
- \hat a(t_{21})^{-1} \hat b(t_{21}) | \phi_1 \rangle
\nonumber \\ & = &
|  \phi_1 \rangle
+ t_{21} \hat \lambda \left[ \hat a_0' + \hat b_0' \right]
|  \phi_1 \rangle
+ | t_{21} | \hat \lambda \left[ \hat a_0 + \hat b_0 \right]
|  \phi_1 \rangle
%\nonumber \\ & & \mbox{ }
%+ {\cal O}(t_{21}^2)
\nonumber \\ & = &
|  \phi_1 \rangle
+ \frac{ t_{21} }{i\hbar}  \hat{\cal H}(t)  \, | \phi_1 \rangle
+ \frac{| t_{21} |}{2} \hat \lambda(t) \,
\hat {\overline S}_\mathrm{r}\!'(t) \, |  \phi_1 \rangle .
\end{eqnarray}
The neglected terms are ${\cal O}(t_{21}^2)$.

From this one can identify one form for the most likely
time propagator, namely
\begin{equation}\label{Eq:hat-U-ne}
\hat{\overline{\cal U}}_\mathrm{ne}(t_2,t_1)
=
\hat{\mathrm I}
+ \frac{ t_{21} }{i\hbar} \hat{\cal H}(t)
+ \frac{| t_{21} |}{2}
\hat \lambda(t) \, \hat {\overline S}_\mathrm{r}\!\!'(t) .
\end{equation}
This gives the conditional most likely wave state  as
\begin{eqnarray}
| \overline \psi(t_2|\psi_1,t_1) \rangle
& = &
\hat{\overline{\cal U}}_\mathrm{ne}(t_2,t_1) \, | \psi_1 \rangle
\nonumber \\  & = &
|  \psi_1 \rangle
+ \frac{ t_{21} }{i\hbar}
\hat{\cal H}(t)  \, | \psi_1 \rangle
\nonumber  \\ &  & \mbox{ }
+ \frac{| t_{21} |}{2}
\hat \lambda(t) \, \hat {\overline S}_\mathrm{r}\!\!'(t)
 |  \psi_1 \rangle ,
%\nonumber  \\ &  & \mbox{ }
%+ \frac{t_{21}-| t_{21} |}{2}
%\hat \lambda(t) \, \hat {\overline S}_\mathrm{r}\!\!'(t)  \,
%\hat{\cal P}_0(t) \,|  \psi_1 \rangle ,
\end{eqnarray}
and the evolution of the ground state projection as
\begin{eqnarray}
| \overline \psi(t_2) \rangle
& = &
\hat{\overline{\cal U}}_\mathrm{ne}(t_2,t_1) \,
| \overline \psi(t_1) \rangle
\\ \nonumber & = &
| \overline \psi(t_1) \rangle
+ \frac{ t_{21} }{i\hbar}
\hat{\cal H}(t)  \, | \overline \psi(t_1) \rangle .
%\nonumber  \\ &  & \mbox{ }
%+ \frac{ t_{21} }{2}
%\hat \lambda(t) \, \hat {\overline S}_\mathrm{r}\!\!'(t)\,
%| \overline \psi(t_1) \rangle .
\end{eqnarray}
The evolution of the ground state projection is purely adiabatic
because the thermodynamic force %, the reservoir entropy gradient,
vanishes in the ground state, Eq.~(\ref{Eq:Sr'psi=Sr''phi}),
$\hat {\overline S}_\mathrm{r}\!'(t)  \, | \overline \psi(t_1) \rangle
=\hat {\overline S}_\mathrm{r}\!'(t)  \, \hat{\cal P}_0(t) \,|  \psi_1 \rangle
= | 0 \rangle$.
In general the propagator, Eq.~(\ref{Eq:hat-U-ne}),
does not couple the sub-system ground state to the reservoir.

Alternatively, using the results of \S \ref{Sec:Sst},
the thermodynamic force can be replaced by its static part.
Writing the fluctuation explicitly in terms of wave states,
Eq.~(\ref{Eq:olphi2}) becomes
\begin{eqnarray}  \label{Eq:olpsi2}
\lefteqn{
| \overline \psi(t_2|\psi_1,t_1) \rangle
- | \overline \psi(t_2) \rangle
}\nonumber  \\
\nonumber \\
& = &
|  \psi_1 \rangle
+ \frac{ t_{21} }{i\hbar}
\hat{\cal H}(t)  \, | \psi_1 \rangle
+ \frac{| t_{21} |}{2}
\hat \lambda(t) \,
\hat {\overline S}_\mathrm{st}\!\!'(t)  \,
\hat{\cal P}_\perp(t) \, |  \psi_1 \rangle
\nonumber \\ && \mbox{ }
- | \overline \psi(t_1) \rangle
- \frac{ t_{21} }{i\hbar}
\hat{\cal H}(t)  \, | \overline \psi(t_1) \rangle ,
%\nonumber \\ && \mbox{ }
%- \frac{| t_{21} |}{2}
%\hat \lambda(t) \, \hat {\overline S}_\mathrm{st}\!\!'(t) \,
%\hat{\cal P}_0(t) \, |  \psi_1 \rangle .
\end{eqnarray}
where the excited state projection is
$ \hat{\cal P}_\perp(t) \equiv \hat{\mathrm I} - \hat{\cal P}_0(t) $.
This expresses the thermodynamic force
in terms of the static part of the entropy fluctuation operator,
Eq.~(\ref{Eq:Sr'-Sst'}),
$ \hat {\overline S}\,'\!\!_\mathrm{r}(t) |  \, \psi_1 \rangle
= \hat {\overline S}\,'\!\!_\mathrm{st}(t)
 \hat{\cal P}_\perp(t) \, |  \psi_1 \rangle $.
Since the starting point
of the most likely trajectory was chosen as
$ | \overline \psi(t_1) \rangle = \hat{\cal P}_0(t_1) \, |  \psi_1 \rangle$,
and since the end point was
$| \overline \psi(t_2) \rangle =
\hat{\overline{\cal U}}(t_2,t_1) \, | \overline \psi(t_1) \rangle$,
one can identify from this the most likely time propagator
\begin{equation} %\label{Eq:hat-U}
\hat{\overline{\cal U}}(t_2,t_1)
=
\hat{\mathrm I}
+ \frac{ t_{21} }{i\hbar} \hat{\cal H}(t)
+ \frac{| t_{21} |}{2}
\hat \lambda(t) \,
\hat {\overline S}_\mathrm{st}\!\!'(t) \, \hat{\cal P}_\perp(t) .
%\nonumber \\ && \mbox{ }
%- \frac{ |t_{21}| }{2}
%\hat \lambda(t) \,\hat {\overline S}_\mathrm{st}\!\!'(t) \,
%\hat{\cal P}_0(t)   .
\end{equation}
This gives the most likely wave state following the transition as
\begin{eqnarray}
| \overline \psi(t_2|\psi_1,t_1) \rangle
& = &
\hat{\overline{\cal U}}(t_2,t_1) \, | \psi_1 \rangle
 \\ \nonumber & = &
|  \psi_1 \rangle
+ \frac{ t_{21} }{i\hbar}
\hat{\cal H}(t)  \, | \psi_1 \rangle
 \\ \nonumber && \mbox{ }
+ \frac{ |t_{21}| }{2}
\hat \lambda(t) \,\hat {\overline S}_\mathrm{st}\!\!'(t) \,
\hat{\cal P}_\perp(t)  \, | \psi_1 \rangle ,
\end{eqnarray}
and the evolution of the ground state projection as
\begin{eqnarray}
| \overline \psi(t_2) \rangle
& = &
\hat{\overline{\cal U}}(t_2,t_1) \,
| \overline \psi(t_1) \rangle
\nonumber \\  & = &
| \overline \psi(t_1) \rangle
+ \frac{ t_{21} }{i\hbar}
\hat{\cal H}(t)  \, | \overline \psi(t_1) \rangle .
%\nonumber  \\ &  & \mbox{ }
%+ \frac{ t_{21} }{2}
%\hat \lambda(t) \, \hat {\overline S}_\mathrm{r}\!\!'(t)\,
%| \overline \psi(t_1) \rangle .
\end{eqnarray}
Again, in general there is no contribution from the ground state
to the reservoir-induced evolution.
There may be contributions from the excited states
to the reservoir-induced evolution of the ground state.
Also, the Hamiltonian operator for the adiabatic evolution
may mix ground and excited states.

The trajectory due to the present dissipative propagator,
$\hat{\overline{\cal U}}_\mathrm{ne}(t_2,t_1) \approx
\hat{\overline{\cal U}}(t_2,t_1)$,
corrects the corresponding  classical\cite{NETDSM,Attard14}
and equilibrium quantum\cite{QSM2} cases
in that here there is no reversible reservoir contribution
(i.e.\ here there is no reservoir term proportional to $t_{21}$).
In earlier work
the fluctuations were from `the' most likely trajectory,
which was argued to be reversible.\cite{NETDSM,Attard14,QSM2}
Here the fluctuations are from the ground state projection,
and reversibility does not come into it.
A further argument for preferring the present formulation
is that the derivation of the adiabatic part of the evolution,
Eq.~(\ref{Eq:a0'+b0'=^H}),
neglected any reversible reservoir contribution contained in the
term $\hat\lambda(t)[ \hat a_0'(t) + \hat b_0'(t)]$
on the grounds that it was small compared with the reversible
adiabatic term itself.
It would therefore be a little inconsistent to invoke
a reversible reservoir contribution for the evolution of the ground state.
Upon reflection, there is no compelling reason
to invoke a reversible trajectory
or to have different propagators for the ground and for the
excited states.
In the light of the present results,
the analysis in Refs.~[\onlinecite{NETDSM,Attard14,QSM2}]
needs to be revisited.

The most likely propagator may be written
\begin{equation}
\hat{\overline{\cal U}}(t_2,t_1)
\equiv
\hat{\mathrm I}
+
\frac{t_{21}}{i\hbar} \hat {\cal H}(t)
+
\hat{\overline {\cal R}}(t_{21},t)
+ {\cal O}(t_{21}^2) ,
\end{equation}
with the dissipative reservoir operator being defined as
\begin{eqnarray} \label{Eq:olR}
\hat{\overline {\cal R}}(t_{21},t)
& \equiv &
\frac{| t_{21} |}{2} \hat \lambda(t) \,
\hat {\overline S}\,'\!\!_\mathrm{st}(t) \hat{\cal P}_\perp(t).
\end{eqnarray}

The static entropy force operator
$\hat {\overline S}\,'\!\!_\mathrm{st}(t)\,$
depends upon the magnitude of the entropy ground state projection
of the wave function, $N(\overline \psi(t))$.
In this sense the most likely propagator
and the dissipative  reservoir operator
are non-linear operators.
However, since this is a constant factor,
it at worst re-scales the thermodynamic force,
and so the non-linearity is  weak and unimportant.
In fact, it could be incorporated in the dissipative operator,
$\hat \lambda(t)$,
which can be arbitrarily scaled by a positive number.

As mentioned in \S \ref{Sec:Sst},
rewriting of the thermodynamic force
in terms of the static part of the thermodynamic force
is effectively the same as replacing
the entropy fluctuation operator
by the static part of the entropy fluctuation operator,
Eq.~(\ref{Eq:Sr''-Sst''}) or (\ref{Eq:Sr'-Sst'}).
This might appear to be an approximation,
but using the expression (\ref{Eq:Sdyn'})
for the dynamic contribution to the force
can be argued to be an exact result.\cite{Attard14}
It appears that the replacement is  necessary on physical grounds,
namely that  in the dissipative Schr\"odinger equation,
the dissipative term comes from the reservoir--sub-system interactions
and it is the static part of the entropy operator
that fully reflects such interactions
(c.f.\ the discussion in the conclusion of Ref.~[\onlinecite{Attard14}]).

One way to see why this replacement is both necessary and exact
is to note the distinction between
the change in entropy and the difference in entropy
(see \S 8.4.1 of Ref.~[\onlinecite{NETDSM}]).
For a transition $\{\psi_1,t_1\} \rightarrow \{\psi_2,t_2\} $,
the difference in entropy is
$S_\mathrm{r}(\psi_2,t_2) - S_\mathrm{r}(\psi_1,t_1)$,
whereas the change in entropy is
$S_\mathrm{st}(\psi_2,t_2) - S_\mathrm{st}(\psi_1,t_1)
- t_{21} \dot S^0_\mathrm{st}(\psi,t)$.
The expression for $S_\mathrm{r}(\psi,t)$
is based upon the average trajectory,
(more precisely, the average of the square of the propagator),
and this is an approximation to the actual entropy
on the specific trajectory being considered in the current transition.
Hence the difference in entropy is an approximation
to the actual change in entropy that occurs in the transition.
Since the static part of the entropy is defined via
the exchange of conserved quantities with the reservoir,
it is exact for such an exchange.
Hence this expression for the change in entropy
that involves the static part of the entropy
gives exactly the change in entropy for the above transition.
This is fundamentally the reason that the above expression for
the dissipative Schr\"odinger equation in terms of the static
part of the thermodynamic force is exact.

%%%%%%%%%%%%%%%%%%%%%%%%%%%%%%%%%%%%%%%%%%%%%%%%%%%%%%%%%%%%%
\subsubsection{Alternatively}

One can incorporate the adiabatic development
into the most likely state about which the final terminus fluctuates.
That is,
choose the reversible reservoir contribution to vanish,
\begin{equation}
t_{21}\hat \lambda \left[ \hat a_0' + \hat b_0' \right]
=
0 ,
\end{equation}
and let
 $| \overline \psi(t_1) \rangle = \hat{\cal P}_0(t) |  \psi_1\rangle$,
and
$| \overline \psi(t_2) \rangle =
| \overline \psi(t_1) \rangle  +
 (t_{21} /i\hbar) \hat{\cal H}(t) | \psi_1\rangle$,
so that the fluctuations are
\begin{eqnarray}
| \phi_1 \rangle
& \equiv &
%\hat{\cal P}_\perp(t_1) | \psi_1 \rangle ,
| \psi_1 \rangle - \hat{\cal P}_0(t_1) | \psi_1 \rangle ,
 \\ \nonumber
\mbox{ and } | \phi_2 \rangle
& \equiv &
| \psi_2 \rangle
- \hat{\cal P}_0(t_1) | \psi_1 \rangle
- \frac{t_{21} }{i\hbar} \hat{\cal H}(t_1) | \psi_1\rangle .
\end{eqnarray}
Then one has
\begin{eqnarray}
\lefteqn{
| \overline \psi(t_2|\psi_1,t_1 \rangle
}  \\ \nonumber
& = &
| \overline \psi(t_2) \rangle
- \hat a(t_{21})^{-1} \hat b(t_{21}) | \phi_1 \rangle
\nonumber \\ & = &
| \overline \psi(t_2) \rangle
+ |  \phi_1 \rangle
+ | t_{21} | \hat \lambda \left[ \hat a_0 + \hat b_0 \right]
|  \phi_1 \rangle
+ {\cal O}(t_{21}^2)
%\nonumber \\ & & \mbox{ }
\nonumber \\ & = &
|  \psi_1 \rangle
+ \frac{ t_{21} }{i\hbar}  \hat{\cal H}(t)  \, | \psi_1 \rangle
%\nonumber \\ & & \mbox{ }
+ \frac{| t_{21} |}{2} \hat \lambda(t) \,
\hat {\overline S}_\mathrm{r}\!'(t) \, |  \psi_1 \rangle .
%+ {\cal O}(t_{21}^2) .
\nonumber \\ & = &
|  \psi_1 \rangle
+ \frac{ t_{21} }{i\hbar}  \hat{\cal H}(t)  \, | \psi_1 \rangle
%\nonumber \\ & & \mbox{ }
+ \frac{| t_{21} |}{2} \hat \lambda(t) \,
\hat {\overline S}_\mathrm{st}\!'(t) \,\hat{\cal P}_\perp(t_1)
\,  |  \psi_1 \rangle . \nonumber
%+ {\cal O}(t_{21}^2) .
\end{eqnarray}
Although the final result is the same as the above,
the physical interpretation is better because the fluctuations
should be the random contribution from the reservoir.

%%%%%%%%%%%%%%%%%%%%%%%%%%%%%%%%%%%%%%%%%%%%%%%%%%%%%%%%%%%%%
\subsection{Stochastic, Dissipative Schr\"odinger Equation} \label{Sec:DSE}

Since the evolution of the sub-system wave function
is determined in part by the interactions with the reservoir,
and since the wave function of the reservoir is indeterminate,
there must be a random element to the evolution.
This means that  the evolution is only determined in a probabilistic sense;
each time that the sub-system visits a particular sub-system wave state
the subsequent evolution is not exactly the same.
This random reservoir propagator acts on the sub-system
and it can be decomposed into the average (or most likely) part,
and the stochastic (or fluctuation) part,
\begin{equation}
\hat{{\cal R}}(t_{21},t)
=
\hat{\overline {\cal R}}(t_{21},t)
+ \hat{\tilde{\cal R}}(t_{21},t) .
\end{equation}
The most likely part is the dissipative reservoir operator given above.
The stochastic operator obviously has  zero mean,
$\langle \hat{\tilde{\cal R}}(t_{21},t) \rangle_\mathrm{stoch} =  0 $.
Adding this stochastic reservoir contribution
to the above deterministic equation
gives  the stochastic, dissipative Schr\"odinger equation,
\begin{eqnarray} \label{Eq:SDSE}
| \psi(t_2) \rangle
& = &
\left[ \hat{\mathrm I}
+
\frac{t_{21}}{i\hbar} \hat {\cal H}(t)
+ \hat{\overline {\cal R}}(t_{21},t)
%\right. \nonumber \\ &  & \left. \mbox{ }
+ \hat{\tilde{\cal R}}(t_{21},t)
\right] | \psi(t_1)\rangle
%+ {\cal O}(t_{21}^2)
\nonumber \\ & \equiv &
\left[ \hat{\overline{\cal U}}(t_{2},t_1)) + \hat{\tilde{\cal R}}(t_{21},t)
\right]
\left| \psi(t_1) \right>
\nonumber \\ & \equiv &
\hat{\cal U}(t_2,t_1) \left| \psi(t_1) \right>.
\end{eqnarray}
The first neglected term here is ${\cal O}(t_{21}^2)$.
The stochastic operator $\hat{\tilde{\cal R}}(t_{21},t)$
and the dissipative (or drag) operator $\hat \lambda(t) $
must be such that  the unitary condition, Eq.~(\ref{Eq:wp1-uni}),
and possibly also the stationarity condition,  Eq.~(\ref{Eq:wp1-stat}),
are satisfied.

The unitary condition,
Eq.~(\ref{Eq:wp1-uni}),
to linear order in the time step is explicitly
\begin{eqnarray}
\hat{\mathrm I}
& = &
\left< \hat{\cal U}(t_2,t_1)^\dag \,
\hat{\cal U}(t_2,t_1)\right>_\mathrm{stoch}
\nonumber \\ & = &
\hat{\mathrm I}
%+ \frac{t_{21}}{i\hbar} \hat {\cal H}(t)
%- \frac{t_{21}}{i\hbar} \hat {\cal H}(t)
%\nonumber \\ &  & \mbox{ }
+
\frac{| t_{21} |}{2}
\left[ \hat \lambda(t) \, \hat {\overline S}\,'\!\!_\mathrm{st}(t)
\hat{\cal P}_\perp(t)
+  \hat{\cal P}_\perp(t)\,
\hat {\overline S}\,'\!\!_\mathrm{st}(t) \, \hat \lambda(t) \right]
\nonumber \\ &  & \mbox{ }
+
\left< \hat{\tilde{\cal R}}(t_{21},t)^\dag \,
\hat{\tilde{\cal R}}(t_{21},t)\right>_\mathrm{stoch} .
\end{eqnarray}
This uses the fact that $\hat \lambda(t)$,
$\hat {\overline S}\,'\!\!_\mathrm{st}(t) $,
and $ \hat{\cal P}_\perp(t)$  are Hermitian.
The adiabatic terms have canceled.
This gives for the variance
\begin{eqnarray} \label{Eq:RR-st}
\lefteqn{
\left< \hat{\tilde{\cal R}}(t_{21},t)^\dag \,
\hat{\tilde{\cal R}}(t_{21},t)\right>_\mathrm{stoch}
}  \\
& = &
\frac{-| t_{21} |}{2}
\left[ \hat \lambda(t) \, \hat {\overline S}\,'\!\!_\mathrm{st}(t)
\,  \hat{\cal P}_\perp(t)
+  \hat{\cal P}_\perp(t) \,
\hat {\overline S}\,'\!\!_\mathrm{st}(t) \,  \hat \lambda(t)
\right] .
 \nonumber
\end{eqnarray}
Hence if the real, symmetric, dissipative operator $\hat \lambda(t)$
is specified, the probability distribution for the random
operator is given by this.
This is the fundamental quantum fluctuation-dissipation theorem.

The stationarity condition, Eq.~(\ref{Eq:wp1-stat}),
to linear order in the time step, is explicitly
\begin{eqnarray}
\hat \wp(t_2)
& = &
\left< \hat{\cal U}(t_1,t_2)^\dag \, \hat \wp(t_1) \,
\hat{\cal U}(t_1,t_2)\right>_\mathrm{stoch}
\nonumber \\ & = &
\hat \wp(t_1)
+ \frac{t_{21}}{i\hbar} \left[
\hat {\cal H}(t) \, \hat \wp(t_1)
- \hat \wp(t_1) \, \hat {\cal H}(t) \right]
\nonumber \\ &  & \mbox{ }
+
\frac{| t_{21} |}{2}
\left[ \hat \wp(t_1) \,
\hat \lambda(t) \, \hat {\overline S}\,'\!\!_\mathrm{st}(t)
\hat{\cal P}_\perp(t)
\right. \nonumber \\ &  & \left. \mbox{ }
+  \hat{\cal P}_\perp(t)\,
\hat {\overline S}\,'\!\!_\mathrm{st}(t) \, \hat \lambda(t)
\hat \wp(t_1) \right]
\nonumber \\ &  & \mbox{ }
+
\left< \hat{\tilde{\cal R}}(t_{12},t)^\dag \, \hat \wp(t_1) \,
\hat{\tilde{\cal R}}(t_{12},t)\right>_\mathrm{stoch} .
\end{eqnarray}
This also gives the evolution of a non-optimum (transient, or approximate)
probability operator.

%\newpage
%%%%%%%%%%%%%%%%%%%%%%%%%%%%%%%%%%%%%%%%%%%%%%%%%%%%%%%%%%%%%%%%%%%%%%%%%%%%%%%
\subsubsection{Ansatz for the Reservoir Operators}
\label{Sec:Ansatz}

The drag and stochastic operators represent the perturbative
interactions of the sub-system with the reservoir.
As such they can be freely chosen, provided that they
satisfy the unitary condition.
%, and %(possibly but not necessarily)\cite{QSM2}
%also the stationary condition, Eq.~(\ref{Eq:wp1-stat}).
%It seems worthwhile to explore tentatively
%two relatively simple ansatz for them
%that seem reasonable on physical grounds.
The simplest ansatz involves a single drag coefficient
and a single stochastic coefficient.
One can take
\begin{equation}
\hat \lambda(t) = -\lambda(t) \,
\hat{\cal P}_\perp(t) \, \hat {\overline S}\,'\!\!_\mathrm{st}(t)
\, \hat{\cal P}_\perp(t),
\end{equation}
and
\begin{equation}
\hat{\tilde{\cal R}}(t) = r(t) \,
\hat{\cal P}_\perp(t) \, \hat {\overline S}\,'\!\!_\mathrm{st}(t)
\, \hat{\cal P}_\perp(t) ,
\end{equation}
with $\lambda(t)$ a positive real number,
and $r(t) $ a real random variable.
(Recall that $\hat \lambda(t)$ is a real, symmetric,
positive semi-definite operator,
and that  $ \hat {\overline S}\,'\!\!_\mathrm{st}(t)$ is a real, symmetric,
and at least approximately negative semi-definite operator.)
The excited state projectors here
confine the drag operator and the stochastic operator to the excited sub-space.
The reason for doing this is that
the operators in the fluctuation form for the second entropy,
Eq.~(\ref{Eq:S2-flucn}), act on the excited states.

Inserting these into the unitary condition (\ref{Eq:RR-st})
one obtains for the variance
\begin{equation}
\left< r(t)^2 \right>_\mathrm{stoch}
=
|t_{21} | \lambda(t)
\end{equation}
This the simplest form of the fluctuation-dissipation theorem.
It says that the variance of the fluctuations
is proportional to the drag coefficient
and to the time step.

%There appear to be no problems for wave states
% $\langle \psi | \hat {\overline S}\,'\!\!_\mathrm{st}(t)
% |\psi \rangle > 0 $,
%so the possible violation of the negativity condition
%may not be serious in practice.
%In any case one could ensure negativity by approximating
%the ground state eigenvalue that appears in the operator
%by the static entropy ground state eigenvalue,
%$S_0(t) \approx S_0^\mathrm{st}(t)$,
%although the full consequences of such an approximation are unclear.

The non-linearity of the thermodynamic force operator
appears as a prefactor that is the magnitude
of the ground state projection of the wave function.
The non-linearity could be removed by
effectively incorporating it into  the arbitrary drag coefficient,
which is the same as setting $N(\overline \psi) =1$ everywhere.

With the above ansatz in the single step time propagator,
the reservoir only couples excited states;
during a transition
it does not influence, nor is it influenced by
the ground state.
However, it is likely the adiabatic term
mixes ground and excited states.

It should be stressed that this ansatz is offered
simply as a possibility worth considering.
It has not been tested in practice
and the full consequences of using it have not been explored.

% end comment %%%%%%%%%%%%%%%%%%%%%%%%%%%%%%%%%%%%%%%%%%%%%%%%%%%%%%%%%

\comment{ %%%%%%%%%%%%%%%%%%%%%%%%%%%%%%%%%%%%%%%%%%%%%%%%%%%%%%
Try the ansatz
$\hat \lambda(t) = \lambda(t) \hat {\overline S}_\mathrm{st}\!\!\!''(t)$
and
$\hat{\tilde{\cal R}}(t_{21},t)
= r(t) \hat {\overline S}_\mathrm{st}\!\!\!''(t)$.
Then
\begin{equation}
%\lefteqn{
\left< r(t)^* \, r(t) \right>_\mathrm{stoch}
% \hat {\overline S}_\mathrm{st}\!\!''(t) \,
%  \hat {\overline S}_\mathrm{st}\!\!''(t)
%} \nonumber \\
=
-| t_{21} | \lambda(t) .
%\hat {\overline S}_\mathrm{st}\!\!''(t) \,
%\hat {\overline S}_\mathrm{st}\!\!''(t)  .
\end{equation}
I can't see anything wrong with this.
} % end comment %%%%%%%%%%%%%%%%%%%%%%%%%%%%%%%%%%%%%%%%%%%%%%%%%%%%%%%%%

%\newpage
%%%%%%%%%%%%%%%%%%%%%%%%%%%%%%%%%%%%%%%%%%%%%%%%%%%%%%%%%%%%%%%%%%%%%%%%%%%%%%%
\subsection{Density Matrix and Statistical Average}

Statistical averages for the non-equilibrium system
may be obtained using the stochastic, dissipative Schr\"odinger equation
by constructing the density matrix for the current wave state.
Conventionally, a statistical average is obtained from a density matrix
that corresponds to a mixture of wave functions,
each one a pure quantum state
resulting from the collapse of the wave function.
\cite{QSM1,Messiah61,Merzbacher70,Bogulbov82}
This is the so-called ensemble approach to statistical mechanics,
which is deprecated by the present author.
In general the density matrix constructed from a single wave function contains
a superposition of states,
and so one cannot use it because such superposition states
should not contribute to the statistical average.

In Paper II,\cite{QSM2}
the canonical equilibrium quantum system was analyzed,
and the corresponding the stochastic, dissipative Schr\"odinger equation
was used to obtain the statistical average
by expressing it as the time average over the trajectory
of the density matrix for the current wave state.
(It is actually the trace of the product of the density matrix and
the operator that occurs.)
The reason that this works is that the phase factors
of the entropy states are randomly distributed
and so averaged  over time
the superposition states cancel from the density matrix
leaving the equivalent of a mixture of pure entropy states
as the only non-zero contributions.

One can do something similar in the present non-equilibrium case,
except of course that
since one wants the average of an operator at a specific time,
and since in general the system is time dependent,
one cannot take a time average over the trajectory.
(For the case of a steady state non-equilibrium system,
such as steady heat flow,
the sub-system does not change macroscopically with time
and it would be possible to take a time average over the trajectory.)
Instead of a time average one can construct an average over multiple
trajectories.

Let $\psi_a(t)$, $a = 1, 2, \ldots, M$,
be the wave state at time $t$
given by the $a$th realization
of the stochastic, dissipative Schr\"odinger equation.
If the trajectories are long enough, $t- t_0 \agt \tau_\mathrm{relax}$,
then the starting wave state is unimportant.
%(assuming stability ---see below).
One could use the same wave state to start each trajectory,
or one could distribute the starting wave states randomly
according to the exact or to an approximate probability distribution.
%perhaps according to the static entropy,
%$\hat \wp_\mathrm{st}(t_0) =
%Z_\mathrm{st}(t_0)^{-1} e^{\hat S_\mathrm{st}(t_0)/k_\mathrm{B} }$.
The  density matrix for the $a$th trajectory is
$\hat \rho_a(t) \equiv | \psi_a(t)\rangle \, \langle \psi_a(t)|
/ N( \psi_a(t) ) $,
and the statistical average of an operator at time $t$ is
\begin{eqnarray}
\left< \hat O(t) \right>_\mathrm{stat}
& = &
\frac{1}{M} \sum_{a=1}^M
\mathrm{TR} \left\{ \hat \rho_a(t) \hat O(t) \right\}
\nonumber \\ & = &
\frac{1}{M} \sum_{a=1}^M
\frac{\langle \psi_a(t)| \hat O(t) | \psi_a(t)\rangle
}{\langle \psi_a(t)| \psi_a(t)\rangle } .
\end{eqnarray}
Although this appears to include superposition states,
only pure entropy quantum states contribute to this
when the average is taken
provided that the magnitudes of the entropy wave states
have converged to a value independent of the initial value,
and the phase factors of the entropy states are uncorrelated.

For example, in terms of the eigenstates of the entropy operator,
$ \hat S_\mathrm{r}(t) | \zeta^S_{\alpha h}(t) \rangle
= S_\alpha(t) | \zeta^S_{\alpha h}(t) \rangle$,
suppose that for  $t- t_0 \agt \tau_\mathrm{relax}$,
the wave function for the $a$th trajectory has representation
\begin{equation}
| \psi_a(t) \rangle
= \sum_{\alpha h} | \psi_{\alpha h}^S(t) |
\, e^{i\theta_{a;\alpha h}(t)} \,
| \zeta^S_{\alpha h}(t) \rangle .
\end{equation}
That is, the amplitude is independent of the particular trajectory.
The entropy representation is special in this regard.
With this and assuming all wave functions have the same normalization,
the average of the expectation value is explicitly
\begin{eqnarray}
%\left< \hat O(t) \right>_\mathrm{stat}
\lefteqn{
\frac{1}{M } \sum_{a=1}^M
\langle \psi_a(t)| \hat O(t) | \psi_a(t)\rangle
} \nonumber \\
& = &
\frac{1}{M } \sum_{a=1}^M
\sum_{\alpha,h} \sum_{\beta,g}
| \psi_{\alpha h}^S(t) | | \psi_{\beta g}^S(t) |
\nonumber \\ & & \mbox{ } \times
e^{-i[\theta_{a;\alpha h}(t)-\theta_{a;\beta g}(t)]}
O_{\alpha h,\beta g}^S(t)
\nonumber \\ & = &
\sum_{\alpha,h}
| \psi_{\alpha h}^S(t) |^2 \,
%\nonumber \\ & & \mbox{ } \times
O_{\alpha h,\alpha h}^S(t) .
\end{eqnarray}
The random phase factors have ensured that the off-diagonal terms
in the entropy representation
have averaged to zero,
\begin{equation}
\frac{1}{M} \sum_{a=1}^M
e^{-i[\theta_{a;\alpha h}(t)-\theta_{a;\beta g}(t)]}
= \delta_{\alpha\beta}\, \delta_{gh} .
\end{equation}
This leaves an expression that is equivalent
to a mixture of pure entropy states.
Provided that the distribution of the magnitudes
converges to the representation of the probability operator,
$| \psi_{\alpha h}^S(t) |^2 \propto \wp_{\alpha,\alpha}^S(t)$,
then this is equal to the statistical average
that would be obtained directly from the probability operator.
One expects that this would be the case
if the stochastic, dissipative propagator satisfies
the stationarity condition, Eq.~(\ref{Eq:wp1-stat}),
since it is then plausible that the non-equilibrium probability operator
is stable under the action of the propagator.

%\newpage
%%%%%%%%%%%%%%%%%%%%%%%%%%%%%%%%
\subsection{Dynamic Part of the Entropy Operator}

%For the present quantum case,
It will be recalled that the probability operator
for the non-equilibrium system
was the exponential of the reservoir entropy operator,
which was the sum of static and dynamic parts,  Eq.~(\ref{Eq:wp(t)}),
$\hat S_\mathrm{r}(t) =
\hat S_\mathrm{st}(t) + \hat S_\mathrm{dyn}(t) $.
The dynamic part of the entropy operator
was given by Eq.~(\ref{Eq:^Sdyn})
\begin{equation}
\hat S_\mathrm{dyn}(t)
=
- \int_0^t \mathrm{d} t' \; \left<
\hat{{\cal U}}(t',t)^\dag
\hat{\dot S}\,\!^0_\mathrm{st}(t')
\hat{{\cal U}}(t',t)
\right>_\mathrm{stoch} .
\end{equation}
Because $t \ge t'$, here
$\hat{\cal U}(t',t)$ is the backward stochastic, dissipative propagator.

In the classical non-equilibrium case,
it was shown that the dynamic part of the entropy could
be replaced by its odd parity projection,
and that the stochastic, dissipative backward trajectories that appeared
could be replaced by adiabatic backward trajectories.
\cite{NETDSM,Attard14}
In the present non-equilibrium quantum case,
one may similarly argue
that the even parity projection of the entropy operator
is dominated by the static entropy operator, which is of course real.
Since the feature that distinguishes a non-equilibrium system
 from its equilibrium counterpart
is that the probability operator cannot be real,
it follows that one must retain,
and that one need only retain,
the odd parity projection of the dynamic entropy operator,
\begin{equation}
\hat S_\mathrm{r}(t) \approx
\hat S_\mathrm{st}(t)
+ \hat S_\mathrm{dyn}^\mathrm{odd}(t) .
\end{equation}
The odd parity projection is
\begin{eqnarray}
\hat S_\mathrm{dyn}^\mathrm{odd}(t)
& \equiv &
\frac{1}{2} \left[
\hat S_\mathrm{dyn}(t) - \hat S_\mathrm{dyn}(t)^* \right]
\nonumber \\ & = &
\frac{-1}{2}
 \int_0^t \mathrm{d} t' \;
 \left[ \left<
\hat{{\cal U}}(t',t)^\dag
\hat{\dot S}\,\!^0_\mathrm{st}(t')
\hat{{\cal U}}(t',t)
\right>_\mathrm{stoch}
\right. \nonumber \\ & & \left. \mbox{ }
- \left<
\hat{{\cal U}}(t',t)^\mathrm{T}
\hat{\dot S}\,\!^0_\mathrm{st}(t')^*
\hat{{\cal U}}(t',t)^*
\right>_\mathrm{stoch}
\right] .
\end{eqnarray}
Following closely the corresponding classical analysis,
(see \S IV of Ref.~[\onlinecite{Attard14}]),
this may be approximated by adiabatic trajectories
\begin{eqnarray}
\hat S_\mathrm{dyn}^\mathrm{odd}(t)
& = &
\frac{-1}{2}  \int_0^t \mathrm{d} t' \;
\left< \hat{{\cal U}}(t',t)^\dag
\hat{\dot S}\,\!^0_\mathrm{st}(t')
\hat{{\cal U}}(t',t)
\right. \nonumber \\ & & \left. \mbox{ }
-  \hat{{\cal U}}(t',t)^\mathrm{T}
\hat{\dot S}\,\!^0_\mathrm{st}(t')^*
\hat{{\cal U}}(t',t)^* \right>_\mathrm{stoch}
\nonumber \\  & \approx &
\frac{-1}{2}  \int_t^{2t} \mathrm{d} t' \;
\left< \hat{{\cal U}}(t',t)^\dag
\hat{\dot S}\,\!^0_\mathrm{st}(t')
\hat{{\cal U}}(t',t)
\right. \nonumber \\ & & \left. \mbox{ }
- \hat{{\cal U}}(t',t)^\mathrm{T}
\hat{\dot S}\,\!^0_\mathrm{st}(t')^*
\hat{{\cal U}}(t',t)^* \right>_\mathrm{stoch}
\nonumber \\  & \approx &
\frac{-1}{2}  \int_t^{2t} \mathrm{d} t' \;
\left< \hat{{\cal U}}^0(t',t)^\dag
\hat{\dot S}\,\!^0_\mathrm{st}(t')
\hat{{\cal U}}^0(t',t)
\right. \nonumber \\ & & \left. \mbox{ }
- \hat{{\cal U}}^0(t',t)^\mathrm{T}
\hat{\dot S}\,\!^0_\mathrm{st}(t')^*
\hat{{\cal U}}^0(t',t)^* \right>_\mathrm{stoch}
\nonumber \\  & = &
\frac{-1}{2}  \int_0^{t} \mathrm{d} t' \;
\left< \hat{{\cal U}}^0(t',t)
\hat{\dot S}\,\!^0_\mathrm{st}(t')
\hat{{\cal U}}^0(t',t)^\dag
\right. \nonumber \\ & & \left. \mbox{ }
- \hat{{\cal U}}^0(t',t)^\dag
\hat{\dot S}\,\!^0_\mathrm{st}(t')^*
\hat{{\cal U}}^0(t',t)  \right>_\mathrm{stoch}   .
\end{eqnarray}
%The first equality is the definition.
The second equality transforms from backward trajectories
to future trajectories,
and is justified because the evolution of the fluctuation in the dissipation
is to a good approximation even in time.
(The difference between the dissipations
may be written as the difference of the difference between each dissipation
and the average dissipation, which is the fluctuation.)
The third equality is  essentially Onsager's regression hypothesis:
\cite{Onsager31a}
the future regression of a fluctuation is the same
in an open system as in an adiabatic or isolated system.
The fourth equality invokes the time reversibility of
the adiabatic propagator.
The adiabatic propagator is symmetric,
$\hat{{\cal U}}^0(t',t)^\mathrm{T}
= \hat{{\cal U}}^0(t',t)$,
and has the time symmetry
$ \hat{{\cal U}}^0(t',t)^* = \hat{{\cal U}}^0(2t-t',t)$.
(For a mechanical non-equilibrium system,
the Hamiltonian operator is extended into the future,
$\hat{\tilde{\cal H}}(t';t) = \hat{\cal H}(2t-t')$, $t'>t$,
in order to preserve this symmetry.
Of course the final equality refers only to $t'<t$
and so this is not needed explicitly.)

One can go further for two common cases.
For a thermodynamic steady state system,
the adiabatic rate of change of the static part of the reservoir
entropy has odd parity,
$\hat{\dot S}\,\!^0_\mathrm{st}(t')^*
= - \hat{\dot S}\,\!^0_\mathrm{st}(t')$.
For a mechanical non-equilibrium system (time-dependent Hamiltonian operator)
it has even parity,
$\hat{\dot S}\,\!^0_\mathrm{st}(t')^*
= \hat{\dot S}\,\!^0_\mathrm{st}(t')$.
Accordingly, one can  define
\begin{eqnarray}
\hat S_\mathrm{dyn}^\mathrm{odd;0}(t)
& \equiv &
\frac{-1}{2}
 \int_0^t \mathrm{d} t' \;
 \left[ \left<
\hat{{\cal U}}^0(t',t)^\dag
\hat{\dot S}\,\!^0_\mathrm{st}(t')
\hat{{\cal U}}^0(t',t)
\right>_\mathrm{stoch}
\right. \nonumber \\ & & \left. \mbox{ }
- \left<
\hat{{\cal U}}^0(t',t)^\mathrm{T}
\hat{\dot S}\,\!^0_\mathrm{st}(t')^*
\hat{{\cal U}}^0(t',t)^*
\right>_\mathrm{stoch}
\right] .
\end{eqnarray}
With this and the above one sees that
\begin{equation}
\hat S_\mathrm{dyn}^\mathrm{odd}(t)
\approx
\pm \hat S_\mathrm{dyn}^\mathrm{odd;0}(t) ,
\end{equation}
with the positive sign for a steady state thermodynamic system,
and the negative sign for a mechanical non-equilibrium system.
Although this has no computational advantages
over the preceding expression,
and although it does not apply to a non-steady state thermodynamic system,
it does have a certain aesthetic appeal
for these two common cases.

With this or else the preceding expression,
the entropy operator is in a form suitable for computation:
because the expression invokes the adiabatic propagator,
one does not have to compute the difficult stochastic, dissipative
backward propagator.

%The adiabatic  total time derivative of an operator is
%\begin{eqnarray}
%\hat{\dot O}\,\!^0(t) & \equiv &
%\partial_t \hat O(t)
%+ \hat{\dot{{\cal U}}}\,\!^0(t)^\dag \hat O(t)
%+ \hat O(t) \hat{\dot{{\cal U}}}\,\!^0(t)
%\nonumber \\ & = &
%\partial_t \hat O(t)
%- \frac{1}{i\hbar} \hat{\cal H}(t) \hat O(t)
%+ \frac{1}{i\hbar} \hat O(t)  \hat{\cal H}(t).
%\end{eqnarray}

%%%%%%%%%%%%%%%%%%%%%%%%%%%%%%%%%%%%%%%%%%%%%%%%%%%%%%%%%%%%%%%%%%%%%%%%%%%%%%%
%                                                                             %
                \section{Executive Summary}
%                                                                             %
%%%%%%%%%%%%%%%%%%%%%%%%%%%%%%%%%%%%%%%%%%%%%%%%%%%%%%%%%%%%%%%%%%%%%%%%%%%%%%%

In this paper two results have been derived from first principles
that lie at the foundations of non-equilibrium quantum statistical mechanics.
First, the non-equilibrium probability operator was obtained,
Eq.~(\ref{Eq:wp(t)}).
This was written as the exponential of an entropy operator,
which in turn was shown to be the sum of a static and a dynamic part.
The static part has the same form as the instantaneous
equilibrium entropy operator,
and the dynamic part is correction that accounts
for the prior adiabatic changes
that occur in non-equilibrium systems.
The form of the probability operator is quite general,
and it applies to mechanical non-equilibrium systems
(i.e.\ time varying potentials),
and to thermodynamic  non-equilibrium systems
(e.g.\ heat flow, hydrodynamic fluxes, chemical reactions).

Second,  the  stochastic, dissipative Schr\"odinger equation
for an open non-equilibrium system was obtained,
Eq.~(\ref{Eq:SDSE}).
Again the formulation is sufficiently general
as to encompass both mechanical and thermodynamic systems.
The time propagator was shown to comprise
adiabatic, dissipative, and stochastic terms.
The dissipative term
was derived from the thermodynamic force
and entropy fluctuation operators,
which are in general non-linear operators.
The variance of the stochastic operator was related to the
dissipative operator by imposing the unitary condition
on the propagator.
This is the  quantum version
of the fluctuation-dissipation theorem.
The non-equilibrium statistical average of an operator at a particular time
was expressed in terms of the average density matrix
generated by the stochastic, dissipative Schr\"odinger equation.

%\newpage $\;$ \newpage
%%%%%%%%%%%%%%%%%%%%%%%%%%%%%%%%%%%%%%%%%%%%%%%%%%%%%%%%%%%%%%%%%%%%%%%%%%

%%%%%%%%%%%%%%%%%%%%%%%%%%%%%%%%%%%%%%%%%%%%%%%%%%%%%%%%%%%%%%%%%%%%%%%%%%

\begin{thebibliography}{99}



\bibitem{AttardV}
Attard, P. (2006),
%``Statistical Mechanical Theory for Steady State Systems.
%V. Non-equilibrium Probability Density.'',
J. Chem.\ Phys.\ {\bf 124}, 224103.

\bibitem{AttardIX}
Attard, P. (2009),
%``Statistical Mechanical Theory for Non-equilibrium Systems. IX.
%Stochastic Molecular Dynamics.''
J. Chem.\ Phys.\ {\bf 130}, 194113.

\bibitem{NETDSM}
Attard, P. (2012),
\emph{Non-Equilibrium Thermodynamics and Statistical Mechanics:
Foundations and Applications},
(Oxford University Press, Oxford).
%ISBN  978-0-19-966276-0


\bibitem{Attard14}
Attard, P. (2014),
%'Simplified Derivation of the Non-Equilibrium Probability Distribution',
arXiv:1405.1469.

\bibitem{Attard00}
Attard, P. (2000),
%``The Explicit Density Functional and its Connection with
%Entropy Maximisation'',
J. Stat.\ Phys.\ {\bf 100}, 445. %--473

\bibitem{TDSM}
Attard, P. (2002),
\emph{Thermodynamics and Statistical Mechanics:
Equilibrium by Entropy Maximisation},
(Academic Press, London).
%ISBN  0-12-066321-X.  Library of Congress 2002103392


\bibitem{QSM1}
Attard, P. (2013),
%`Quantum Statistical Mechanics. I.
%Decoherence, Wave function Collapse, and the von Neumann Density Matrix',
arXiv:1401.1786.


\bibitem{QSM2}
Attard, P. (2013),
%`Quantum Statistical Mechanics. II.
%Stochastic Schr\"odinger Equation',
arXiv:1401.1787v3.

\bibitem{QSM3}
Attard, P. (2014),
%'Quantum Statistical Mechanics. III. Equilibrium Probability',
arXiv:1404.2683.
%In this paper the fluctuation expansions
%were performed on the expectation entropy,
%which is a linearization of the correct actual entropy.


%\bibitem{Neumann27}
%von Neumann, J. (1927),
%%"Wahrscheinlichkeitstheoretischer Aufbau der Quantenmechanik",
%G\"ottinger Nachrichten {\bf 1}, 245. %–272.

\bibitem{Messiah61}
Messiah, A. (1961),
\emph{Quantum Mechanics},
(North-Holland, Amsterdam, Vols I and II).


\bibitem{Merzbacher70}
Merzbacher, E. (1970),
\emph{Quantum Mechanics},
(Wiley, New York, 2nd edn).

\bibitem{Bogulbov82}
Bogulbov, N. N. and Bogulbov, N. N. (1982),
\emph{Introduction to Quantum Statistical Mechanics},
(World Scientific, Singapore).

%\bibitem{Gasiorowicz74}
%Gasiorowicz, S. (1974), \emph{Quantum Physics}, (Wiley, New York).


%%%%%%%%%%%%%%%%%%%%%%%%%%%%%%




%%%%%%%%%%%%%%%%%%%%%%%%%%%%%%%%%%%%%%%%%%%%%%%%%%%%
% Formulation of non-linear quantum mechanics:
% non-linear quantum mechanics:

%Peter NATTERMANN Symmetry in Nonlinear Mathematical Physics 1997, V.2, 270–278.
% http://www.imath.kiev.ua/~appmath/Symmetry97/art36.pdf

%Contrary to these, we are concerned with a more fundamental role of
%nonlinearity in quantum mechanics. Notable efforts in this direction
%have been launched, for example, by Bialinycki-Birula and Mycielski
%[4], Weinberg [5], and Doebner and Goldin [6, 7]

\bibitem{Birula76} %[4]
Bialynicki-Birula, I. and Mycielski, J. (1976)
Ann.\ Phys.\ (NY), {\bf 100}, 62. %-–93.

\bibitem{Weinberg89} %[5]
Weinberg S. (1989),
Phys.\ Rev.\ Lett.\ , {\bf 62}, 485. %-–488.

\bibitem{Doebner92} %[6]
Doebner, H.-D. and Goldin, G. A. (1992),
Phys.\ Lett.\ A, {\bf 162}, 397. %-–401.


\bibitem{Doebner94} %[7]
Doebner, H.-D. and Goldin, G. A. (1994),
J.\ Phys.\ A: Math.\ Gen.\ {\bf 27}, 1771. %-–1780.


\bibitem{Doebner95} % [32]
Doebner, H.-D., Dobrev, V. K., and Nattermann, P., (1995),
(editors),
\emph{Nonlinear, Deformed and Irreversible Quantum Systems},
(World Scientific, Singapore).

%%%%%%%%%%%%%
\bibitem{Bugajski91}
Bugajski, S. (1991),
Int.\ J. Theor.\ Phys.\ {\bf 30}, 961. %-971
%Nonlinear quantum mechanics is a classical theory

\bibitem{Beretta87}
Beretta, G. P., (1987),
%Quantum Thermodynamics of Nonequilibrium
%Onsager Reciprocacy and Disperison Dissipation Relations
Foundations of Physics {\bf 17}, 365.
%see also www.ing.unibs.it/~beretta/www.quantumthermodynamics.org/WebSite1.pdf





%%%%%%%%%%%%%%%%%%%%%%%%%%%%%%%%%%%%%%%%%%%%%%%%%%%%
% non-linear Schr\"odinger equation:

%In view of the intensive studies on nonlinear
%Schr\"odinger equations in the last decade, it
%is astonishing to note that a framework for a consistent framework
%of nonlinear quantum theories has already been given by Mielnik in 1974 [13]


%\bibitem{Mielnik74} %[13]
%Mielnik, B. (1974),
%Comm.\ Math.\ Phys.\ {\bf 37}, 221. %–256


%Although the studies of the non-linear modifications of the Schrödinger
%equation are not new [4]-[14]
\bibitem{deBroglie60} %[4]
de Broglie, L. (1960),
\emph{Nonlinear Wavemechanics}, (Elsevier, Amsterdam).

\bibitem{Laurent65} %[5]
Laurent, B.  and Roos, M.  (1965),
Nuovo Cimento {\bf 40}, 788.

\bibitem{Shapiro73} %[6]
Shapiro, I. R.  (1973),
Sov.\ J.\ Nucl.\ Phys.\ {\bf 16}, 727.

\bibitem{Marinov74} %[7]
Marinov, M. S.  (1974),
Sov.\ J.\ Nucl.\ Phys.\ {\bf 19}, 173.

\bibitem{Kupczynski74} %[8]
Kupczynski, M.  (1974),
Lett.\ Nuovo Cimento {\bf 9}, 134. %, sr 2 no 4, 134.


\bibitem{Mielnik74} %[9] (also [13] above in Natterman
Mielnik, B. (1974),
Comm.\ Math.\ Phys.\ {\bf 37}, 221. %–256

\bibitem{Pearle76}%[10]
Pearle, P.  (1976),
Phys.\ Rev.\ D {\bf 13}, 857.

%\bibitem{Birula76} %[11]
%I. Bialinicki-Birula and J. Mycielski, Ann. Phys. 100, 62 (1976).

\bibitem{Shimony79} %[12]
Shimony, A. (1979),
Phys.\ Rev.\ A {\bf 20}, 394.

\bibitem{Kibble78} %[13]
Kibble, T. W. B. (1978),
Comm.\ Math.\ Phys.\ {\bf 64}, 73 (1978); {\bf 65}, 189 (1979).

\bibitem{Kibble80} %[14]
Kibble, T. W. B.  and Randjbar-Daemi, S. (1980),
J.\ Phys.\ A {\bf 13}, 141.

%%%%%%%%%%%%%%%%%%%%%%%%%%%%%%%%%%%%%%%%%%%%%%%%%%%%
% non-linear Schr\"odinger equation + friction & dissipation:

%There have also been attempts to incorporate friction
%on a microscopic level using non-linear Schr\"odinger equations.
%Many of these approaches incorporate stochastic frictional
%forces in the nonlinear evolution equation for wavefunctions (see, e.g., [3]).
\bibitem{Messer78} %[3]
Messer J. (1978),
Lett.\ Math.\ Phys.\ {\bf 2}, 281. %–-286.

%%%%%%%%%%%%%%%%%%%%%%%%%%%%%%%%%%%%%%%%%

\bibitem{Moxnes05}
Moxnes, J. F. and Hausken, K.  (2005),
Annales de la Fondation Louis de Broglie,  {\bf 30}, 309.
%'A Non-Linear Schrödinger Equation Used to Describe Friction'

\bibitem{Lange85}
Lange H. (1985),
%`On Nonlinear Schrodinger Equations In
%The Theory Of Quantum Mechanical Dissipative Systems'
Nonlinear Analysis: Theory, Methods and Applications
{\bf 9}, 1115.%-1133.

\bibitem{Garashchuk13}
Garashchuk, S., Dixit, V., Gu, B., and Mazzuca, J., (2013),
J.\ Chem.\ Phys.\  138:054107. %doi: 10.1063/1.4788832.
%The Schro?dinger equation with friction from the quantum trajectory perspective.

%the nonlinear Schr\"odinger{Langevin equation is taken into granted
%to describe the dissipative process due to frictional force, for
%instance, in the motion of a Brownian particle in heat bath by
%Kostin [15], to characterize directly a class of nonlinear quantum
%mechanics through nonlinear gauge generaliza- tion by
%Doebner-Goldin-Nattermann [5], and to study the motion of charged
%(quantum) particles in semiconductor of nano-size [12, 13]

\bibitem{Kostin72} %[15]
Kostin, M. D. (1972),
%'On the Schrodinger{Langevin equations,'
J.\ Chem.\ Phys.\ {\bf 57}, 2589. %-3591(?).

\bibitem{Doebner99} %[5]
Doebner, H.-D., Goldin, G.A., and Nattermann, P. (1999),
%'Gauge transformations in quantum mechanics
%and the unification of nonlinear Schrodinger equations'
J.\ Math.\ Phys.\ {\bf 40},  49.%-63.

\bibitem{Weiss93} %[25]
Weiss, U.  (1993),
\emph{Quantum Dissipative Systems},
(World Scientific, Singapore).

\bibitem{Dekker81} %[4]
Dekker, H. (1981),
%'Classical and quantum mechanics of damped harmonic oscillator,
Phys.\ Rep.\ {\bf 80}, 1. %--112

\bibitem{Schuch12}
Schuch, D. (2012),
J.\ Phys.\ Conf.\ Ser.\ {\bf 380},  012009
%'Complex Riccati equations as a link between different approaches
%for the description of dissipative and irreversible systems'



%%%%%%%%%%%%%%%%%%%%%%%%%%%%%%%%%%%%%%%%%%%%%%%%%%%%%

\bibitem{AttardII}
Attard, P. (2005),
% ``Statistical Mechanical Theory for Steady State Systems.
%II. Reciprocal Relations and the Second Entropy.''
J. Chem.\ Phys.\ {\bf 122}, 154101.



\bibitem{AttardI}
Attard, P. (2004),
%``Statistical Mechanical Theory for the Structure of Steady State Systems.
%Application to a Lennard-Jones Fluid with Applied Temperature Gradient'',
J.\ Chem.\ Phys.\ {\bf 121}, 7076. %--7085 (2004)


\bibitem{AttardIII}
Attard, P. (2005),
%``Statistical Mechanical Theory for Steady State Systems.
%III. Heat Flow in a Lennard-Jones Fluid.''
J. Chem.\ Phys.\ {\bf 122}, 244105. % (2005)


\bibitem{Onsager31a}
Onsager, L. (1931)
Phys.\ Rev. {\bf 37}, 405.


\bibitem{Green54}
Green, M. S. (1954),
J. Chem.\ Phys.\ {\bf 22}, 398.

\bibitem{Kubo66}
Kubo, R. (1966),
Rep.\ Progr.\ Phys.\ {\bf 29}, 255.




\bibitem{AttardVIII}
Attard, P. and Gray-Weale, A. (2008),
%``Statistical Mechanical Theory for Steady State Systems. VIII.
%General Theory for a Brownian Particle
%Driven by a Time- and Space-Varying Force.'',
J. Chem.\ Phys.\  {\bf 128}, 114509.  %[2]

\bibitem{Attard09}
Attard, P. (2009),
%``Non-equilibrium Monte Carlo Simulation for a Driven Brownian Particle'',
Phys.\ Rev.\ E {\bf 80}, 041126.



\end{thebibliography}
\end{document}